\documentclass[a4paper,11pt]{article}
\pdfoutput=1 

\usepackage{jheppub} 
\usepackage[utf8]{inputenc}
\usepackage{pdfrender}

\usepackage{xcolor}  
\usepackage{pdflscape}
\usepackage[caption=false,font=footnotesize,labelformat=simple]{subfig}
\usepackage{amsmath, bm}
\usepackage[utf8]{inputenc}

\title{\boldmath  Improvements on perturbative oscillation formulas 
including non-standard neutrino Interactions 
}

\author[a,b,c]{M. E. Chaves,}
\author[d]{D. R. Gratieri,\note{Corresponding author.}}
\author[c,1]{O. L .G. Peres}

\affiliation[a]{Instituto de F\'isica - UFF, {24210-310}, Niter\'oi RJ, Brazil}
\affiliation[b]{Instituto de Ci\^encias Exatas - UFF, {27213-145}, Volta Redonda RJ, Brazil}
\affiliation[c]{Instituto de F\'isica Gleb Wataghin - UNICAMP, {13083-859}, Campinas SP, Brazil}
\affiliation[d]{Escola de Engenharia Industrial  Metal\'urgica de Volta Redonda - UFF, {27225-125}, Volta Redonda RJ, Brazil}

\emailAdd{mchaves@ifi.unicamp.br, ORCID:0000-0001-7396-081X}
\emailAdd{drgratieri@id.uff.br,ORCID:0000-0002-4561-5805}
\emailAdd{orlando@ifi.unicamp.br,  ORCID:0000-0003-2104-8460}

\abstract{
 We use perturbation theory to obtain neutrino oscillation probabilities, including the standard mass-mixing paradigm and non-standard neutrino
 interactions (NSI). The perturbation made  in standard parameters $\Delta m^2_{21}/\Delta m^2_{31}$ and $\sin^2{(\theta_{13})}$ and in the non-diagonal NSI parameters,
 but keep diagonal NSI parameters non-perturbated. We perform the calculation for the channels $\nu_{\mu}\to \nu_{e}$ and $\nu_{\mu}\to \nu_{\mu}$.  The resulting 
 oscillation formulas are compact and present functional structure similar to the standard oscillation  (SO) case. They apply to a wide range in the allowed NSI space
 of parameters and include the previous results from perturbative approaches as limit cases. We also take advantage of have compact formulas to {\it explain} the origin of the degeneracies in the neutrino probabilities in terms of the invariance of amplitude and phase of oscillations. Then we determine analytically the multiple sets of combinations of SO and NSI parameters that result in oscillation probabilities identical to the SO case. 
}

\begin{document} 
\maketitle

\section{Introduction}
\label{Sec:intro}

 The neutrino mass-mixing formalism that emerged from the last decades of neutrino 
 phenomenology~ \cite{fukuda1998evidence,anselmann1994gallex,abdurashitov1994results,fukuda2001constraints,Ahmad:2002jz,apollonio2003search,Eguchi:2002dm,an2012observation,adamson2011measurement,allison1999atmospheric,ambrosio1998measurement,Aartsen:2019eht,T2K,cite-key}
 is actually  known as {\it Standard Neutrino Oscillations} (SO), and contains 6 different parameters.  Two mass differences $(\Delta m^{2}_{31}, \Delta m^{2}_{21})$,
 one CP phase $(\delta_{\rm CP})$ and three mixing angles, ($\theta_{12}, \theta_{13}, \theta_{23}$).  The current  values for these parameters can be found in the 
 Ref.~\cite{Esteban:2018azc}.
 As a consequence of SO, 
 the difference between the squared  neutrino mass eigenvalues must be  $ \Delta m^{2}_{3l} \approx 2.523 \times 10^{-3}$ eV$^{2}$ and  $ \Delta m^{2}_{21}=7.39 \times 10^{-5}$ eV$^{2}$, where l mean lighter neutrinos.
 However, in the {\it Standard Model of Particles and Fields} (SM) from the Refs.  \cite{PhysRevD.2.1285,Weinberg:1967tq,salam1968elementary} and  others,  
 neutrinos are included as massless particles. This suggests that the mechanism responsible to give mass to the neutrino should be other than the Brout-Englert-Higgs model \cite{Englert:1964et,Higgs:1964ia,Higgs:1964pj} and the experimental discovery in 
 Refs.~\cite{Atlas-Higgs,CMS-Higgs}.  Henceforth, it is straightforward to search for physics beyond the SM to account for neutrino masses.  In this sense,
 in this work we take into account the so-called {\it Non-Standard Neutrino Interactions} (NSI) which was firstly proposed by Wolfenstein \cite{wolfenstein1978neutrino,wolfenstein1979neutrino}. NSI  can give strong resonant effects even for unmixed neutrinos~\cite{guzzo1991msw,Guzzo:1991cp,Roulet:1991sm,valle1987resonant} and also for massless neutrinos~\cite{valle1987resonant}.
 In the standard three-neutrino  formalism,  the time evolution of a neutrino flavor state $\{|\nu_{e},\nu_{\mu},\nu_{\tau}, \rangle \}$  is given by a
 Schroedinger-like equation.  When  NSI are taken into account~\cite{Asano:2011nj,ohlsson2013status,miranda2015non,Esteban:2019lfo}, the neutrino time evolution equation assumes the form:
\begin{eqnarray}
i\dfrac{d}{dt} \begin{pmatrix}  \nu_{e} \\ \nu_{\mu}\\ \nu_{\tau} \end{pmatrix}  &=& H\begin{pmatrix}  \nu_{e} \\ \nu_{\mu}\\ \nu_{\tau}  \end{pmatrix}=\left({\cal H}+H_{\rm NSI}\right)\begin{pmatrix}  \nu_{e} \\ \nu_{\mu}\\ \nu_{\tau}  \end{pmatrix}~,  
\label{Eq:Schr01}
\end{eqnarray}
where 
%
\begin{eqnarray}
{\cal H}= \Delta_{31} \left[U\begin{pmatrix}
0 & 0 & 0 \\
0 & r_{\Delta} & 0 \\
0 & 0 & 1
\end{pmatrix}U^\dagger +r_{A}\begin{pmatrix}
1 & 0 & 0 \\
0 & 0 & 0 \\
0 & 0 & 0
\end{pmatrix}\right],
\label{Eq:Schr02}
\end{eqnarray}
and the NSI Hamiltonian,  with the complex non-diagonal
NSI parameters, and the real diagonal NSI parameters,
\begin{eqnarray}
\quad H_{\rm NSI}=\Delta_{31} r_{A}  \epsilon^{\prime}= \Delta_{31} r_{A}\begin{pmatrix}
\epsilon_{\rm ee}^{\prime}& \epsilon_{e\mu}^{\prime} & \epsilon_{e\tau}^{\prime} \\
 \epsilon_{\mu e}^{\prime}& \epsilon_{\mu\mu}^{\prime} & \epsilon_{\mu\tau}^{\prime} \\
 \epsilon_{\tau e }^{\prime} & \epsilon_{\tau\mu}^{\prime} & \epsilon_{\tau\tau}^{\prime}
\end{pmatrix}.
\label{Eq:Schr02a}
\end{eqnarray}
We adopt the  parametrization  for the mixing matrix,  $U={\rm R} (\theta_{23}) {\rm \widetilde{R}} (\theta_{13},\delta_{\rm CP}) {\rm R}(\theta_{12})$~\cite{patrignani2016review}. Here 
${\rm R}(\theta_{ij})$ is a  rotation of the angle $\theta_{ij}$ in the  {\it i-j} plane and ${\rm \widetilde{R}} (\theta_{13},\delta_{\rm CP})$ is a complex rotation by an angle $\theta_{13}$ and  phase $\delta_{\rm CP}$.
To keep the notation short, we define, 
\begin{equation}
\Delta_{31}=\Delta m^2_{31}/2E_{\nu}~,~~~~r_\Delta=\dfrac{\Delta m^2_{21}}{\Delta m^2_{31}}~,~~~~r_A=\dfrac{A}{\Delta m^2_{31}}~,
\label{Eq:parex}
\end{equation}
where $\Delta m_{ji}^2 \equiv m^{2}_{j}-m^{2}_{i}$ is the mass square difference between the two mass eigenstates {\em j} and {\em i}. Also, $A=2E_{\nu}V_{\rm CC}$ and $V_{\rm CC}=\sqrt{2}G_{\rm F}n_{e}$ is the matter potential that neutrinos feel while they cross a medium  with the electron number density $n_{e}=N_{A}\rho\langle Z/A \rangle$. $N_{A}$ is
the number of  Avogadro, $\rho$ is the matter density, and $\langle Z/A \rangle$ is the averaged ratio between the nuclear charge and mass number in the medium that neutrino crosses. In this work 
we assume the following values for the mixing angles,  $\sin^2 \theta_{12}=0.31, \sin^2 \theta_{13}=0.023,\sin^2 \theta_{23}=0.5$,
and squared mass differences, $\Delta m^2_{31}=2.4\times 10^{-3}$~eV$^2$, $\Delta m_{21}^2=7.5\times 10^{-5}$~eV$^2$. In the last term in Eq.~(\ref{Eq:Schr01}) we include the effective NSI parameters, $\epsilon_{\alpha \beta}^{\prime}$, as effective matter potentials summed up in medium element in the neutrino
time evolution  Hamiltonian   and which are related  to the complex coupling constants $\epsilon^{\rm f\, V}_{\alpha\beta}$ by,
\begin{equation}
\epsilon_{\alpha\beta}^{\prime}=\sum_{\rm f=e,u,d} Y_f(x)
\epsilon^{\rm f\, V}_{\alpha\beta}, ~~~~~~\alpha, \beta = e, \mu, \tau.
\label{Eq:epsl}
\end{equation}
Here, $Y_f(x)=n_f(x)/{ n_e(x)}$, where $n_f(x)$ is the number of fermions in the medium and the vector coupling  $\epsilon^{\rm f\, V}_{\alpha\beta}=\epsilon^{\rm f\, L}_{\alpha\beta}+\epsilon^{\rm f\, R}_{\alpha\beta}$
 are the  complex coupling parameters  in the  non-standard interaction effective  Lagrangian
\begin{equation}
-\mathcal{L}^{{\rm eff}}_{{\rm NSI}}=   \sum_{\rm f=e,u,d}\sum_{X=L,R} 
\epsilon^{\rm f\, P}_{\alpha\beta}2\sqrt{2}G_F(\overline{\nu}_\alpha\gamma_\rho L\nu_\beta)(\overline{f}\gamma^\rho X f)+(c.c),
\end{equation}
where  $X=(L,R)=(1-\gamma^5,1+\gamma^5) {/\sqrt{2}}$. At the fundamental level, $\epsilon^{\rm f\, P}_{\alpha\beta} \equiv \frac{G_X}{G_f}$ is related with to couplings of neutrino flavor states with 
the electrons and quarks due the  exchanged boson X and it expresses the ratio between the strength of new interaction, $G_X$,  to the strength of SM weak coupling, $G_f$. 

For the oscillation phenomenology,  we can always remove a global phase,  which we performed by subtracting  a multiple of identity matrix in Eq.~(\ref{Eq:Schr01}). We choose to subtract  $\epsilon_{\mu\mu}^{\prime}\times$-Identity matrix and in this way, the only physical parameters for oscillation it will be $\epsilon_{\tau\tau}^{\prime}-\epsilon_{\mu\mu}^{\prime}$ and $\epsilon_{ee}^{\prime}-\epsilon_{\mu\mu}^{\prime}$. We will redefine  the complex matrix $\epsilon^{\prime} $ in terms of the complex matrix $\epsilon$,
\begin{eqnarray}
\epsilon^{\prime} \to 
 \epsilon\equiv \Delta_{31} r_{A}\begin{pmatrix}
\epsilon_{\rm ee}& \epsilon_{e\mu} & \epsilon_{e\tau} \\
 \epsilon_{\mu e}& \epsilon_{\mu\mu} & \epsilon_{\mu\tau} \\
 \epsilon_{\tau e } & \epsilon_{\tau\mu} & \epsilon_{\tau\tau}
 \end{pmatrix}
 =\Delta_{31}r_A
 \begin{pmatrix}
\epsilon_{\rm ee}^{\prime}-\epsilon_{\mu\mu}^{\prime}& \epsilon_{e\mu}^{\prime} & \epsilon_{e\tau}^{\prime} \\
 \epsilon_{\mu e}^{\prime}& 0 & \epsilon_{\mu\tau}^{\prime} \\
 \epsilon_{\tau e }^{\prime} & \epsilon_{\tau\mu}^{\prime} & \epsilon_{\tau\tau}^{\prime}-\epsilon_{\mu\mu}^{\prime}
\end{pmatrix},
\label{mudanca}
\end{eqnarray}
and we will work for now on, with the variables, $\epsilon_{\alpha\beta}$, that for $\alpha=\beta$ are real parameters and for $\alpha \neq  \beta$ are complex parameters.

Since the first studies on NSI~\cite{wolfenstein1978neutrino,guzzo1991msw,Guzzo:1991cp,Roulet:1991sm,valle1987resonant,Grossman:1995wx,Krastev:1997cp,Brooijmans:1998py,GonzalezGarcia:1998hj,Bergmann:2000gp,Guzzo:2000kx,Guzzo:2001mi}, now we have an intense activity on this topic~\cite{Coloma:2011rq,deGouvea:2019ozk,Esteban:2019lfo,Coloma:2016,Coloma:2017egw,Yasuda:2020cff,Rahman:2015vqa,Girardi:2014kca,Szafron:2012mi,Dutta:2020che,Choubey:2019osj,Farzan:2017xzy,Liao:2016orc,Tang:2017qen,Babu:2020nna,Babu:2019iml,Demidov:2019okm,Capozzi:2019iqn,Feng:2019mno,Mitsuka:2011ty,Altmannshofer:2018xyo,Coloma:2015kiu,Ohlsson:2013ip,Choubey:2015xha,Ge:2016dlx,Barenboim:2018lpo,Pandey:2019apj,Gouvea:2015,Flores:2018kwk,Falkowski:2018dmy,Chatterjee:2018dyd,Bischer:2018zcz,Asano:2011nj,kikuchi2009perturbation,Liao:2016hsa,Deepthi:2016erc,Deepthi:2017gxg,Dey:2018yht,Yasuda:2007jp,Meloni:2009ia,Meloni:2009cg,Kopp:2007ne,Blennow:2016etl,Chatterjee:2014gxa,Masud:2015xva,Masud:2016bvp,Masud:2016gcl,Masud:2018pig} and recent reviews can be found in the Refs.~\cite{ohlsson2013status,miranda2015non,Farzan:2017xzy,Esteban:2019lfo}.
 For instance, in  Ref.~\cite{guzzo1991msw,Guzzo:1991cp} it was proposed as a solution for solar neutrino problem and confirmed in  Ref.~\cite{Bergmann:2000gp}. But  from the results of KamLand experiment~\cite{Eguchi:2002dm}, the only possibility it is  a mixed solution of NSI and mass-mixing as discussed in  Ref.~\cite{Guzzo:2004ue}. Similarly for atmospheric neutrinos, the NSI scenario was able to explain the muon neutrino deficit at lower energy as discussed in  Ref.~\cite{GonzalezGarcia:1998hj} but not the muon neutrino deficit at higher energies  in such way that nowadays only mixed solutions are possible.  Bounds on the NSI parameters were made in the literature under two assumptions,
 \begin{enumerate}
     \item assume one and only one non-zero NSI parameter ~\cite{Asano:2011nj,Meloni:2009cg,Mitsuka:2011ty,Kopp:2007ne}
     \item assume more than one non-zero NSI parameter ~\cite{GonzalezGarcia:1998hj,Bergmann:2000gp,Guzzo:2004ue,kikuchi2009perturbation,Miranda:2004nb,Friedland:2004ah,Coloma:2015kiu,Coloma:2016,Gouvea:2015,Esteban:2018ppq,Blennow:2016etl,Masud:2015xva,Masud:2016bvp,Masud:2016gcl,Liao:2016hsa,Deepthi:2016erc,Deepthi:2017gxg,Dey:2018yht,Chatterjee:2014gxa,Ge:2016dlx,Meloni:2009ia}
 \end{enumerate}
 One example of the second case it was found in solar neutrino analysis, where two good solutions were found.  One solution is compatible with vanishing NSI parameters and mixing angles compatible with the values mentioned above, that have $\theta_{12}<\pi/4$. The other solution, called  LMA-Dark solution, it has  solar mixing angle $\theta_{12}>\pi/4$, which  is possible for large  diagonal real parameters NSI, $\epsilon_{ee}$ and $\epsilon_{\mu\mu}/\epsilon_{\tau\tau}$ ~\cite{Miranda:2004nb}. Another example is the analysis of atmospheric data made in the Ref.~\cite{Friedland:2004ah} where solutions with a large NSI parameter $\epsilon_{ee}$,
 $\epsilon_{\tau\tau}$, and the magnitude of  $\epsilon_{e\tau}$
 are virtually identical to the vacuum oscillation and are called {\it vacuum mimicking} solutions. These two examples are a subset of fundamental degeneracy of neutrino probability, where you can have the same oscillation probability with completely different set of parameters for the mixing parameters and  for the NSI parameters. 
  This fundamental degeneracy  happens in 
  neutrino evolution system,
  Eq.~(\ref{Eq:Schr01}), called
 {\it generalized mass ordering degeneracy}~\cite{Coloma:2016}\footnote{A note of caution that in  Ref.~\cite{Coloma:2016} the  fundamental degeneracy of the oscillation probability it was formulated using a different 
 convention for mixing matrix that it was not equal to more common used convention in  Particle Data definition for mixing matrix~\cite{patrignani2016review}. Our Eq.~(\ref{eq:osc-deg}) it is  for Particle Data Group  convention.}, 
 \begin{equation}
  \label{eq:osc-deg}
    \begin{aligned}
    \Delta m^2_{31} &\to  -\Delta m^2_{32} \,,
    \\
    \theta_{12} & \to \pi/2 - \theta_{12} \,,
    \\
    \delta_{\rm CP} &\to 2\pi - \delta_{\rm CP}, \\
   \epsilon_{\alpha\alpha}
    &\to -2\delta_{\alpha e}-  \epsilon_{\alpha\alpha}   \, ,
    \\
     \epsilon_{\alpha\beta}
    &\to - \epsilon_{\alpha\beta}^*  \, , \alpha,\beta \neq e,
    \\
    \epsilon_{e\alpha}
    &\to  \epsilon_{e\alpha}^* .
  \end{aligned}
\end{equation}

 This degeneracy is exact for neutrinos crossing  a constant density   medium,  as it is in the present long-baseline experiments, which are sensitive to the neutrino oscillation probabilities  resulting form the solution of  the neutrino evolution described in  Eqs.~(\ref{Eq:Schr01}-\ref{Eq:Schr02a}).  To  break this degeneracy, we can use the
fact that the differential cross-section for coherent elastic scattering of a $\nu_{\alpha}$ with energy $E_{\nu}$ and a nucleus N with Z protons, N neutrons, and mass M is given by,  assuming that the complex NSI parameters~\cite{Barranco:2005yy}
\begin{equation}
\dfrac{d\sigma_{\nu_{\alpha}\text{-}\mathcal{N}}}{d T}
(E_{\nu},T)
=
\dfrac{G_{\text{F}}^2 M}{\pi}
\left(
1 - \dfrac{M T}{2 E^2_{\nu}}
\right)
Q_{\alpha}^2
,
\label{cs}
\end{equation}
where $T$ is the nuclear recoil kinetic energy and
\newcommand{\vet}[1]{\ensuremath{\hskip-1pt\vec{\hskip1pt#1}}}
\begin{eqnarray}
Q_{\alpha}^2
=
\null & \null
\left[
\left( g_{V}^{p} + 2 \epsilon_{\alpha\alpha}^{uV} + \epsilon_{\alpha\alpha}^{dV}\
 \right)
Z
F_{Z}(|\vet{q}|^2)
+
\left( g_{V}^{n} + \epsilon_{\alpha\alpha}^{uV} + 2 \epsilon_{\alpha\alpha}^{dV}\
 \right)
N
F_{N}(|\vet{q}|^2)
\right]^2
\nonumber
\\
\null & \null
+
\sum_{\beta\neq\alpha}
\left|
\left( 2 \epsilon_{\alpha\beta}^{uV} + \epsilon_{\alpha\beta}^{dV} \right)
Z
F_{Z}(|\vet{q}|^2)
+
\left( \epsilon_{\alpha\beta}^{uV} + 2 \epsilon_{\alpha\beta}^{dV} \right)
N
F_{N}(|\vet{q}|^2)
\right|^2
,
\label{Qalpha2}
\end{eqnarray}
with
\begin{equation}
g_{V}^{p}
=
\dfrac{1}{2} - 2 \sin^2\!\theta_{W}
,
\qquad
g_{V}^{n}
=
- \dfrac{1}{2}
.
\label{gV}
\end{equation}
Here
$\theta_{W}$ is the weak mixing angle,
given by
$\sin^2\!\theta_{W} = 0.23857 \pm 0.00005$
at low energies~\cite{Tanabashi:2018oca}, $F_{Z}(|\vet{q}|^2)$
and
$F_{N}(|\vet{q}|^2)$
are, respectively, the form factors of the proton and neutron distributions in the nucleus,
that depend on the three-momentum transfer $|\vet{q}| \simeq \sqrt{2 M T}$.

Recently the first measurement of 
coherent neutrino-nucleon scattering experiment was made by the COHERENT experiment~\cite{Akimov:2017ade}. 
Also, there are upper limits for these cross-sections from the CONNIE experiment~\cite{Aguilar-Arevalo:2019jlr,Aguilar-Arevalo:2019zme} that it got until now the best limits for light vector mediator search~\cite{Aguilar-Arevalo:2019zme}. The  coherent neutrino-nucleon scattering measurement allows us to get an independent bound on NSI parameters~\cite{Akimov:2017ade,Liao:2017uzy,Shoemaker:2017lzs,Gonzalez-Garcia:2018dep,Farzan:2018gtr,Denton:2018xmq,AristizabalSierra:2018eqm, Altmannshofer:2018xyo,Coloma:2019mbs,Giunti:2019xpr,Canas:2019fjw}. Using data from the COHERENT experiment~\cite{Akimov:2017ade} we can put bounds on 
$\epsilon_{\alpha\beta}^{uV}$ and $\epsilon_{\alpha\beta}^{dV}$
for $\alpha=e,\mu$ and $\beta~=~e,\mu,\tau$. 

Furthermore,  there is a global analysis of all neutrino oscillation data under the NSI framework ~\cite{Esteban:2018ppq} that also combine it with this recent results from COHERENT experiment. Indeed, in this analysis, the authors assume all NSI parameters are real, and  also subtract a diagonal element $\epsilon_{\mu\mu}$, as we did in Eq.~(\ref{mudanca}), which implies that theirs bounds applies to the parameters $\epsilon_{\alpha\beta}$.  The combination of global analysis of neutrino oscillation data and the data from the COHERENT experiment gives upper bounds on the NSI couplings as follows,
\begin{eqnarray}
\epsilon_{\alpha\beta}(\rm min:\rm max)\le \begin{pmatrix}
-0.65:1.4 & -0.19:0.16 & -1.1:0.43  \\
 -0.19:0.16  & -- & -0.05: 0.04  \\
 -1.1:0.43 & -0.05: 0.04 & -0.025:0.50 
\end{pmatrix}.
\label{eq:limits-coherent}
\end{eqnarray}
 We also acknowledged the table of $\Delta \chi^2\times $ NSI parameters in  Ref.~\cite{test1}. 

From these results we can notice a pattern on the bounds on NSI parameters,
\begin{enumerate}
\item the diagonal NSI parameters can have larger values $\epsilon_{\alpha\alpha} \sim (-0.65 \to 1.4)$,
\item non-diagonal NSI parameters can have at maximum $\epsilon_{\alpha\beta}\simeq  (-1.1 \to 0.43 )$ for $\alpha \neq \beta$,
\item  also for the bounds on the non-diagonal NSI paramaters there is a hierarchy. The bound on $\epsilon_{e\tau}$, is relatively weak compared with the bound on $\epsilon_{e\mu}$ and $\epsilon_{\mu\tau}$. From this we assume, that $\epsilon_{e\tau}$ can be larger than $\epsilon_{e\mu}$ and $\epsilon_{\mu\tau}$. 
\end{enumerate}

This pattern of the bounds and also the results from Refs.~\cite{Friedland:2004ah, Miranda:2004nb}  inspire us to think what can be the effects of {\it significant diagonal NSI elements} in the phenomenology, and it was one of the motivations of this work. Indeed, in the recent analysis of the future experiment's sensitivity to the NSI~\cite{Coloma:2015kiu,Coloma:2016,Gouvea:2015}, large diagonal NSI elements are also allowed. In these works it observes two branches for the allowed regions, the first branch {\it (i)} it is compatible with the scenario of null  NSI, and it has as best fit values all NSI parameters $\epsilon_{\alpha\beta} \sim 0$, and second branch {\it (ii)} it is compatible with a scenario with significant NSI parameters.

The fact that large {\em diagonal} NSI parameters are still allowed it is not widely known. For a first approach what can make the effects of large diagonal NSI element in the neutrino phenomenology, we decide to study the effects in the neutrino propagation. We choose to use the perturbation theory approach, and we can get an analytical formula for the neutrino probability.

 Indeed, after the seminal work from  Ref.~\cite{cervera2000golden}, a plethora of methods to solve the Eq.~(\ref{Eq:Schr01}) based on perturbative strategies emerged in the literature. For the standard case (without NSI) it is used  that some parameters of neutrino time evolution Hamiltonian are today known to be small,  $\sin^{2}(\theta_{13}) \approx \Delta m^{2}_{21}/\Delta m^{2}_{31} \approx 0.03$. See for example \cite{Akhmedov:2004ny,Akhmedov:2001kd,cervera2000golden,Asano:2011nj,kikuchi2009perturbation,Meloni:2009cg,Kopp:2007ne,Blennow:2016etl,Liao:2016hsa,Deepthi:2016erc,Deepthi:2017gxg,Dey:2018yht,Ge:2016dlx,Masud:2015xva,Masud:2016bvp,Masud:2016gcl}. For NSI, also there is the use of the perturbation theory, but due it is still unknown the true strength of NSI parameters and  we have different assumptions made in the literature~\cite{Coloma:2019mbs,Asano:2011nj,kikuchi2009perturbation,Meloni:2009cg,Kopp:2007ne,Blennow:2016etl,Liao:2016hsa,Deepthi:2016erc,Deepthi:2017gxg,Dey:2018yht,Ge:2016dlx,Masud:2015xva,Masud:2016bvp,Masud:2016gcl}.

The paper is organized as follows:   In Section \ref{Sec:mot}, we introduce the perturbative methods, the formalism of neutrino propagation through the quantum perturbation theory of Hamiltonian systems, without NSI,  and in Section \ref{Sec:Smat} with NSI. 
In Section \ref{Sec:res-I}, we present the resulting probabilities from the perturbation method, compare it with numerical solutions, and apply it to the DUNE case. In Section \ref{Sec:degenerate}, we show how the results from perturbation theory can explain the degenerate behavior of neutrino time-evolution. Conclusions are in Section \ref{Sec:Conc}.

\section{Perturbative approaches and the neutrino time-evolution}
\label{Sec:mot}

The NSI formalism, as it is given in Eq.~\eqref{Eq:Schr01}, describes a three-neutrino system with the addition of NSI parameters. Firstly we will discuss the standard case, where we have only the three-neutrino system with standard matter effect and later the case for NSI. In the standard oscillation case, exact solutions of Eq.~\eqref{Eq:Schr01} are possible for 
vacuum \cite{giunti2007fundamentals} as well as for constant matter case \cite{cervera2000golden,Kimura:2002hb,ohlsson2000three}. However, for varying matter potentials, full solutions of the  Schroedinger equation are only possible numerically. Henceforth, a common approach in the literature is to use perturbative methods to found approximate semi-analytical solutions. Such works use the perturbative quantum theory for time-dependent Hamiltonian~\cite{sakurai1995modern}.  
Even the solution for
constant density is not easily known, understandable the physical meaning. A solution is given by
Refs.~\cite{Zaglauer:1988gz,bellandi1997resonances}\footnote{We should be aware that there are mistyping in 
Ref.~\cite{Zaglauer:1988gz}, accordingly
with  Ref.~\cite{denton2016compact}} use the eigenvalues and eigenvectors of the $3\times 3$  matrix that is not much illuminating. Other analytical solutions involve 
\begin{itemize}
\item a perturbative approach of full oscillation probability such as (I) small $\theta_{13}$ perturbative expansion ~\cite{cervera2000golden} 
(II) {\it large} $\theta_{13}$ expansion~\cite{Asano:2011nj, Akhmedov:2004ny}. 
\item  a specific rotation that made the problem separable in two $2\times 2$ systems~\cite{Ioannisian:2018qwl,Ge:2016dlx},
\item diagonalization of neutrino evolution~\cite{Yasuda:2007jp}.
\end{itemize}
In  Ref.~\cite{Akhmedov:2001kd}, three neutrino oscillation probabilities are developed within the perturbation theory for an arbitrary
density profile. Because of the non-zero value of $\theta_{13}$ \cite{an2012observation,ahn2012observation}, 
further developments of this formalism were done to extend the theory~\cite{Asano:2011nj, Akhmedov:2004ny}.  
Therefore, in this work we adopt the expansion  parameters, the variables, $(\sin (\theta_{13}), r_{\Delta},\epsilon_{\alpha\beta})$ for $\alpha \neq \beta$.
The state of the art of perturbative methods applied
to neutrino time evolution was studied in ~\cite{denton2016compact,minakata2016simple}. In a recent analysis in Ref.~\cite{Chatterjee:2018dyd}, perturbation theory is used to
analytically take into account the different density values that neutrinos feel while crossing the pathway for DUNE experiment~\cite{Abi:2018dnh}. 
In  Section \ref{Sec:Smat} we will show a perturbative approach that has applicability of NSI parameters  $\epsilon_{\alpha\beta}$ for larger diagonal NSI elements.

\section{ Perturbation Theory with NSI}
\label{Sec:Smat}

Here we address the formalism to solve neutrino time-evolution, including NSI. We  use perturbation theory through Dyson Series and consider as guidelines that the final expression
for the probabilities should obey the following conditions: 
\begin{enumerate}
\item - To agree with the numerical solution;
\item - Allow us to use  the perturbation theory for all the set of values of NSI parameters given in~ Eq.~(\ref{eq:limits-coherent});
\item - Be concise enough to allow direct interpretation and use;
\item - Have the functional form as close as possible to the standard oscillation case in the presence of matter effects.
\end{enumerate}

In the case of standard solution for neutrino oscillation (the limit of $\epsilon_{\alpha\beta}\to 0$ in Eq.~(\ref{Eq:Schr01})) we have that the standard matter effect it is invariant under $\theta_{23}$ rotation. If we define a propagation basis such as 
\begin{equation}
|\widetilde{\nu}_{\alpha}\rangle =[R(\theta_{23})]^{\dagger} |\nu_{\alpha}\rangle,
\label{Eq:baserot}
\end{equation}
where $R(\theta_{23})$ is the rotation matrix by the mixing angle $\theta_{23}$ as defined previously,
then we have two effects:
\begin{enumerate}
    \item the mixing angle $\theta_{23}$  did not appear the new  evolution equation for neutrinos in the $|\widetilde{\nu}_{\alpha}\rangle$ basis,
    \item the Hamiltonian presents a  more simpler format, having a block-diagonal form.
\end{enumerate}
 Once we solved the neutrino evolution in the $|\widetilde{\nu}_{\alpha}\rangle$ basis we should rotated back to get the solution of neutrino evolution. This made the perturbative form of the evolution matrix for standard neutrino oscillation to be more simpler to write.
When we include the NSI term this invariance under a rotation on $\theta_{23}$ it is broken. However, we can keep a simpler format for Hamiltonian, even in the NSI formalism, where the new Hamiltonian is changed to $H \to  \widetilde{H}$, where 
\begin{equation}
\widetilde{H}=[{\rm R}(\theta_{23})]^{\dagger}H{\rm R}(\theta_{23})=
[{\rm R}(\theta_{23})]^{\dagger}\left(\cal{H}\right){\rm R}(\theta_{23})+
[{\rm R}(\theta_{23})]^{\dagger}\left(H_{\rm NSI}\right){\rm R}(\theta_{23}),
\label{Eq:basisH}
\end{equation}
where the last term 
can be recast in terms of the new NSI parameters $\widetilde{\epsilon}$,
\begin{equation}
[{\rm R}(\theta_{23})]^{\dagger}\left(H_{\rm NSI}\right){\rm R}(\theta_{23})= \Delta_{31}r_A 
 [{\rm R}(\theta_{23})]^{\dagger} \epsilon {\rm R}(\theta_{23})\equiv 
  \Delta_{31}r_A \widetilde{\epsilon}.
  \label{eq:HipIn2}
\end{equation}
Here  the notation $\widetilde{\epsilon}$ will be a shorthand  for the product $[{\rm R}(\theta_{23})]^{\dagger} \epsilon {\rm R}(\theta_{23})$.
For each non-diagonal element $\widetilde{\epsilon}_{\alpha\beta}= |\widetilde{\epsilon}_{\alpha\beta}| e^{i\phi_{\alpha \beta}}$ for all $\alpha \neq \beta = e, \mu, \tau$ and the diagonal elements are always real numbers. The explicit relation between $\epsilon$ and $\widetilde{\epsilon}$ is given in the Appendix~\ref{apa}.

A good way to achieve the conditions 1-4 is to create a hierarchy of NSI parameters in the Hamiltonian.  We will divide our full Hamiltonian $\widetilde{H}$ as given in Eq.~(\ref{Eq:basisH}) as follows. We will define
a non-perturbated Hamiltonian that we will called $\widetilde{H}^{(0)}$ 
\begin{eqnarray}
\widetilde{H}^{(0)}=\Delta_{31}\begin{pmatrix}
r_A & 0 & 0 \\
0 & 0 & 0 \\
0 & 0 & 1 \\
\end{pmatrix}+\Delta_{31} r_A\begin{pmatrix}
\widetilde{\epsilon}_{\rm ee}& 0 & 0 \\
0&  0 & 0 \\
0 & 0 & \widetilde{\epsilon}_{\tau\tau}
\end{pmatrix}.
\label{Eq:Htil0}
\end{eqnarray}
In this Hamiltonian, we incorporate the diagonal NSI parameters, $\widetilde{\epsilon}_{\rm ee}$ and $\widetilde{\epsilon}_{\tau\tau}$ (the other $\widetilde{\epsilon}_{\mu\mu}$ was already subtracted off). We will consider this as the non-perturbated Hamiltonian. In other words, we will made no expansions for the diagonal NSI elements. Here is {\it one of the choices of this work}, to incorporate the diagonal NSI parameters as the non-perturbative parameters. This it will have impact on the better expansion for larger diagonal NSI elements, but still inside the $3\sigma$ allowed values at present from global fits, as it will be clear in  Section ~\ref{Sec:res-I} and in Figure~\ref{fig:mina}.

For the next term, we include it in the first term of perturbation Hamiltonian. We {\it choose to have the $s_{13}$ and we incorporate the NSI parameter $\widetilde{\epsilon}_{e\tau}$}, as follows 
\begin{eqnarray}
\widetilde{H}^{(a)}=\Delta_{31}\begin{pmatrix}
0 & 0 & s_{13}e^{-i\delta_{\rm CP}} \\
0 & 0 & 0 \\
s_{13}e^{i\delta_{\rm CP}} & 0 & 0 \\
\end{pmatrix}+\Delta_{31} r_A\begin{pmatrix}
0& 0 & \widetilde{\epsilon}_{e\tau} \\
0& 0 & 0 \\
\widetilde{\epsilon}_{\tau e} & 0 & 0
\end{pmatrix}.
\label{eq:minakataordemmeio}
\end{eqnarray}

With this choice, that it is twofold, we have a block-diagonal format for the perturbated Hamiltonian, and we assume that this NSI parameter is so  important as the $s_{13}$ term. Notice that both expansion parameters located in the same entry of the Hamiltonian. The remain NSI parameters  {\it we will choose} to be next-next-order parameter  of the perturbation theory, which means that $\widetilde{\epsilon}_{e\mu}$ and $\widetilde{\epsilon}_{\mu\tau}$ are putting  together with the $r_{\Delta}$ parameter.  It follows from {\it our choice} that:
\begin{itemize}
    \item $\widetilde{\epsilon}_{e\mu}$ parameter is assumed to have the same order of magnitude than $r_{\Delta}$ parameter. Also, the parameter  enter in the same  position of the respective SO term in the Hamiltonian
    \item Between the available possibilities, {\it we choose} to assume   $\widetilde{\epsilon}_{\mu\tau}$  with the same order than $r_{\Delta}$, since it is not desirable that the parameter appears only at higher orders of the perturbation theory. 
\end{itemize}

\begin{eqnarray}
\widetilde{H}^{(b)}=\Delta_{31}\begin{pmatrix}
r_\Delta s_{12}^2+s_{13}^2 & r_\Delta c_{12}s_{12} & 0 \\
r_\Delta c_{12}s_{12} & r_{\Delta}c_{12}^2 & 0 \\
0 & 0 & -s^2_{13} \\
\end{pmatrix}+\Delta_{31} r_A\begin{pmatrix}
0& \widetilde{\epsilon}_{e\mu} & 0 \\
\widetilde{\epsilon}_{\mu e}& 0 & \widetilde{\epsilon}_{\mu\tau} \\
0 & \widetilde{\epsilon}_{\tau\mu} & 0
\end{pmatrix},
\label{eq:minakataordemum}
\end{eqnarray}

Summarizing, the leading expansion parameter is $(\sin \theta_{13},\widetilde{\epsilon}_{e\tau})$ and the next-leading parameters are $r_{\Delta},\widetilde{\epsilon}_{e\mu}$ and $\widetilde{\epsilon}_{\mu\tau}$.

To complete, we should include the remaining parts of full Hamiltonian  $\widetilde{H}$ that are not include in $\widetilde{H}^{(0)},\widetilde{H}^{(a)}$ and $\widetilde{H}^{(b)}$, that we split in two Hamiltonian
\begin{eqnarray}
\widetilde{H}^{(c)}=-\Delta_{31}\begin{pmatrix}
0 & 0 & (r_\Delta s_{12}^2+\frac{1}{2}s_{13}^2)s_{13}e^{-i\delta_{\rm CP}} \\
0 & 0 & r_\Delta s_{12}c_{12}s_{13}e^{-i\delta_{\rm CP}} \\
(r_\Delta s_{12}^2+\frac{1}{2}s_{13}^2)s_{13}e^{i\delta_{\rm CP}} & r_\Delta s_{12}c_{12}s_{13}e^{i\delta_{\rm CP}} & 0 \\
\end{pmatrix},
\label{Eq:htil11}
\end{eqnarray}
and
\begin{eqnarray}
\widetilde{H}^{(d)}=-\Delta_{31} r_\Delta\begin{pmatrix}
s_{12}^2s_{13}^2 & \dfrac{1}{2}c_{12}s_{12}s_{13}^2 & 0 \\
\dfrac{1}{2}c_{12}s_{12}s_{13}^2 & 0 & 0 \\
0 & 0 & -s_{12}^2s_{13}^2 \\
\end{pmatrix}.
\label{Eq:htil13}
\end{eqnarray}
Due the different dependence of expansion parameters, one have terms like $s_{13}^3, s_{13}r_{\Delta}$ and other have $s_{13}^2r_{\Delta}$. These Hamiltonians did not have any NSI. 

 Henceforth, our choice of the hierarchy can summarize as 

\begin{enumerate}
    \item the diagonal NSI elements, $\widetilde{\epsilon}_{\alpha\alpha}$, are chosen to be included  in the non-perturbated Hamiltonian;
    \item the complex off-diagonal NSI parameters are to be considered sub-leading and sub-sub-leading;
    \subitem $\bullet$ the NSI coupling $\widetilde{\epsilon}_{e\tau}\sim\sqrt{\kappa}$ is  chosen to be included in first order of perturbated Hamiltonian;
    \subitem $\bullet$ the  off-diagonal couplings  $\widetilde{\epsilon}_{e\mu}\sim \widetilde{\epsilon}_{\mu\tau} \sim \kappa$, are chosen to be included in the high order in the  perturbated Hamiltonian,

\end{enumerate}

The full Hamiltonian $\widetilde{H}$ is given by
\begin{equation}
\widetilde{H} =\widetilde{H}^{(0)}+\widetilde{H}^{(a)}+\widetilde{H}^{(b)}+\widetilde{H}^{(c)}+\widetilde{H}^{(d)},
\label{Eq:Hlinha}
\end{equation}
The neutrino time-evolution Hamiltonian, $\widetilde{H}$, as given in Eqs.~(\ref{Eq:htil13}-\ref{Eq:Hlinha}),  can now be evolved in time using the perturbation theory described in Appendix~\ref{Ap:Dyson}.
 The desired probabilities can be written as 
 \begin{equation}
P (\nu_{\alpha}\rightarrow\nu_{\beta})\equiv 
|S_{\beta\alpha}|^{2},
\end{equation}
where 
the $S$ matrix is the neutrino evolution matrix given explicitly from Eq.~(\ref{S:flavor}). 

When we do this computation, we get an infinite series for the probability. We are going to use {\it our choice of NSI hierarchies} to organize this series. The diagonal NSI elements are included 
in the non-perturbated Hamiltonian. Examples where you can this expression with diagonal NSI elements it are in Refs.~\cite{Liao:2016hsa,Deepthi:2016erc,Deepthi:2017gxg,Dey:2018yht}.
As said before,  the leading expansion parameter is $(s_{13}$ and $\widetilde{\epsilon}_{e\tau})$, and the next-leading parameters are $r_{\Delta},\widetilde{\epsilon}_{e\mu}$ and $\widetilde{\epsilon}_{\mu\tau}$. The first non-zero term that appear for $P(\nu_{\mu}\to \nu_e)$ probability is in first order of expansion (there is no zero-order contribution) depends on  $s_{13}^2$ in the standard neutrino oscillation scenario, that came from the $s_{13}$ dependence in the $\widetilde{H}^{(a)}$. An explicit form for the muon neutrino to electron neutrino probability for the first order of expansion 
it given in the standard scenario in Ref.~\cite{Asano:2011nj}.
Physically speaking, this is related in the standard scenario with the {\it atmospheric oscillation} and the mixing angle $\theta_{13}$ relevant for long-baseline neutrino experiments.
The dependence on $r_{\Delta}$ appear only at higher orders in the Hamiltonian $\widetilde{H}^{(b)}$ that is connected with the so-called {\it solar oscillation} with the mixing angle $\theta_{12}$, and which is subdominant for distances until $~7000$~km. This hierarchy used to have in the perturbative analytical solutions for the standard oscillation scenario. Now coming back to NSI, when we introduce the $\widetilde{\epsilon}_{e\tau}$ parameter in $\widetilde{H}^{(a)}$,  
this assumption means that $\widetilde{\epsilon}_{e\tau}$ and the mixing parameter $s_{13}$ will have the contribution
for the same perturbation order. In other words, we will keep everywhere terms proportional to $s_{13}$ and $\widetilde{\epsilon}_{e\tau}$.  Then, for the first order of perturbation theory, we expect to have 
terms like $|\widetilde{\epsilon}_{e\tau}|^2$ and interference terms like $\widetilde{\epsilon}_{e\tau}\times s_{13}$. The explicit expression for the the muon neutrino to electron neutrino probability is given in 
~\cite{Liao:2016hsa,Deepthi:2016erc,Deepthi:2017gxg,Dey:2018yht,Yasuda:2007jp,Meloni:2009ia,Meloni:2009cg,Kopp:2007ne,Blennow:2016etl,Chatterjee:2014gxa,Masud:2015xva,Masud:2016bvp,Masud:2016gcl}. For the other NSI parameters, our choice to have had them in $\widetilde{H}^{(b)}$ it implies that we only appear in the second order in perturbation theory.  For $\widetilde{\epsilon}_{e\mu}$ the probability is listed in 
Refs. ~\cite{Liao:2016hsa,Deepthi:2016erc,Deepthi:2017gxg,Dey:2018yht,Yasuda:2007jp,Meloni:2009ia,Meloni:2009cg,Kopp:2007ne,Blennow:2016etl,Chatterjee:2014gxa,Masud:2015xva,Masud:2016bvp,Masud:2016gcl}. Finally, for $\widetilde{\epsilon}_{\mu\tau}$ a form for this probability it appears in Ref.~\cite{Meloni:2009ia,Asano:2011nj}.

In resume, to perform the perturbation theory, we notice that from the best fit values of mixing angle $\theta_{13}$ and the ratios of squared mass differences, we have an  hierarchy between $s_{13}$ and $r_{\Delta}$,  such that we can define the parameter $\kappa$ as: 
\begin{equation}
 \sin^{2} \theta_{13}\approx r_{\Delta} \approx 0.03 \to \kappa =0.03.
\label{Eq:orderkappa1}
\end{equation}
Another point is what we call small or more significant for a NSI parameter. We will call the NSI parameters small when they are expected to be  numerically (much) smaller than the unity. From the current bounds given in Eq.~(\ref{eq:limits-coherent}), we can see that this is the case for the parameters $\epsilon_{e\mu} \ll 1$, and  $\epsilon_{\mu\tau} \ll 1$. However,  the constrains over $\epsilon_{e\tau}$  are relatively weaker than the bounds over the other two non-diagonal NSI parameters. In fact, in this case we assume $\epsilon^{2}_{e\tau}\ll 1$. The hierarchy of NSI parameters with respect to the parameter $\kappa$ is then  
\begin{equation}
\kappa\approx \epsilon_{\mu\tau}^{\rm small}   \approx \epsilon_{e\mu}^{\rm small}\approx (\epsilon^{2}_{e\tau})^{ \rm small}.
\end{equation}

The explicit  transformation of the NSI parameters from one basis to the other is given in Eq.~(\ref{Eq:epslbasis}).  Furthermore, it is possible to see from Table~\ref{tab:basis} that Eq.~(\ref{Eq:epslbasis}) does not alter the hierarchy between the limits on the NSI parameters when we change from a basis to the other. In the same table we also include  our assumption for the NSI hierarchy (also in the rotated basis $\widetilde{\epsilon}_{\alpha \beta}$) given in the Eqs.~(\ref{Eq:Htil0} - \ref{Eq:htil13}). From a direct inspection,  the last two columns of Table~\ref{tab:basis} it  is possible to conclude that:

\begin{itemize}
\item For the parameters $\widetilde{\epsilon}_{ee}$ and $\widetilde{\epsilon}_{\tau\tau}$, our assumption covers all the allowed domain.
\item Our premise for the hierarchy of the parameter $\widetilde{\epsilon}_{e\tau}$ is that it is in the same order of magnitude than the current experimental bound at $3\sigma$. 
\item For the parameters $\widetilde{\epsilon}_{e\mu}$ and $\widetilde{\epsilon}_{\mu\tau}$ our assumption on the NSI hierarchy is only one order of magnitude below the current limit at $3\sigma$.
\end{itemize}
 As already commented, the bounds from the literature assume that all NSI parameters, $\epsilon$, are real. 
 We can translate to $\epsilon^{\prime}$ notation as defined in Eq.~(\ref{Eq:epsl}) and translated our bounds to $\epsilon^{\prime}_{\alpha\beta}$, if we assume that there is no NSI with electrons: $\epsilon_{\alpha\beta}^{eV}=0$.

\begin{table}[h]
\centering
\begin{tabular}{|c|c|c|c|c|c|c|c|c|c|}
\hline
Parameter &  {\small{$3\sigma$ limit on $\epsilon_{\alpha\beta}\le $ (min:max)}}  & {\small $3\sigma$ limit translated to $\widetilde{\epsilon}_{\alpha\beta}$}  & {\small Our typical  $\widetilde{\epsilon}_{\alpha\beta}$}  \\ 
\hline
$\epsilon_{ee}$         & -0.65 : 1.40 & -0.65 : 1.40 & 1.0 \\
$\epsilon_{\tau\tau}$   & -0.02 : 0.50 & -0.06 : 0.29 & 0.5 \\
$\epsilon_{e\tau}$      &-1.10 : 0.43 & -0.64 : 0.29 & 0.17 \\
$\epsilon_{e\mu}$       & -0.19 : 0.16& -0.14 : 0.46 & 0.03\\
$\epsilon_{\mu\tau}$    & -0.05 : 0.04 & -0.25 : 0.01 & 0.03\\
 \hline
\end{tabular}
\caption{ Comparison between the limits on the NSI parameters, $|\epsilon_{\alpha\beta}|\le $ (min:max) given in  Eq.~(\ref{eq:limits-coherent}), the same limits translated to the basis $\widetilde{\epsilon}_{\alpha \beta}$ through Eq.~(\ref{Eq:epslbasis}) assuming $\theta_{23} = 45^{o}$, and the typical NSI values assumed in this work (also in the rotated basis $\widetilde{\epsilon}_{\alpha \beta}$) given in  Eqs.~(\ref{Eq:Htil0} - \ref{Eq:htil13}).  Here we use all NSI parameters are real.} 
\label{tab:basis}
\end{table}

Henceforth, from the above discussed in Table~\ref{tab:basis} and the associated text, we can expect that our resulting oscillation formulas are applicable in a full region of the current NSI allowed parameter space.

{
\begin{table}[h]
\centering
\begin{tabular}{|c|c|c|c|c|c|c|c|c|c|}
\hline
Parameter &This Work & \cite{Asano:2011nj}&\cite{kikuchi2009perturbation}&     
\cite{Liao:2016hsa,Deepthi:2016erc,Deepthi:2017gxg,Dey:2018yht}
&\cite{Meloni:2009ia}&\cite{Meloni:2009cg,Kopp:2007ne,Blennow:2016etl,Chatterjee:2014gxa}&  \cite{Masud:2015xva,Masud:2016bvp,Masud:2016gcl}\\ 
\hline
$s_{13}$                  & 4 & 4 & 2  & 2  &  2 &  2&2\\ 
$r_{\Delta}$              & 2 & 2 &2 &   2  & 2  & 2 &2 \\ 
 \hline
$\epsilon_{ee}$         & all orders & 1 & --   &  all orders &1  & -- & --\\
%
%
$\epsilon_{\tau\tau}$   & all orders & 1 & --   & -- & 1  & -- & -- \\
$\epsilon_{e\tau}$      & 4 & 2 & 2   &  1  & 1  & 1  &1 \\
$\epsilon_{e\mu}$       & 4 & 2 & 2   &  1 &  1 & 1 &1 \\
$\epsilon_{\mu\tau}$    & 1 & 1 & --  &  -- & 1 & -- &-- \\
\hline
\end{tabular}
\caption{Comparison between the maximum order of the  SO and NSI parameters that appears in our oscillation formulas, as it are given in  Eqs.~(\ref{eq:result1}) and (\ref{Eq:lambdas}),   and in the previous literature. The $"-"$ notation means that the parameter is not present in the given formula.
} 
\label{tab:dist}
\end{table}
}

Moreover, we can characterize the different assumptions about NSI  in two large categories:

\begin{enumerate}
\item include linear dependence of NSI off-diagonal elements~\cite{Masud:2015xva,Masud:2016bvp,Masud:2016gcl,Kopp:2007ne,Blennow:2016etl,Chatterjee:2014gxa,Ge:2016dlx,Meloni:2009cg,Meloni:2009ia},
\item include  NSI off-diagonal elements until second order, this work and  Refs.~\cite{Asano:2011nj,kikuchi2009perturbation,Liao:2016hsa,Deepthi:2016erc,Deepthi:2017gxg,Dey:2018yht}. 
\end{enumerate}
Within these groups, we can have different assumption for the {\it diagonal NSI elements}, as follows: {\it (i)} we can  have non-perturbative diagonal NSI as this work and Ref.~\cite{Ge:2016dlx,Meloni:2009cg,Meloni:2009ia}, {\it (ii)} to have  diagonal NSI parameters in first-order ~\cite{kikuchi2009perturbation,Ge:2016dlx,Meloni:2009cg,Meloni:2009ia,Asano:2011nj}, and {\it (iii)} did not include diagonal NSI effect~\cite{Kopp:2007ne,Blennow:2016etl,Chatterjee:2014gxa,Masud:2015xva,Masud:2016bvp,Masud:2016gcl}. A disclaimer is that these different assumptions are justified in the respective works and inside their context all expansions are correct. What we want to point is that, given the present knowledge of upper bounds on the NSI elements, our assumptions are: to include the real NSI diagonal elements as non-perturbative and include a hierarchy of NSI parameters allow us to cover a wider range of NSI parameters and faithfully reproduce the numerical results of neutrino evolution.  Another point is due to our choice of NSI hierarchy, and we can {\it reproduce} all perturbative models of neutrino probability as case limits of our expression. For example, if we keep only the first order of diagonal NSI elements in our results, we can reproduce 
 Ref.~\cite{Asano:2011nj}.

\section{Results for perturbation theory until second order in expansion parameter, \texorpdfstring{$\kappa^2$}{TEXT}}
\label{Sec:res-I}

At this point  we present our results for the neutrino oscillation probabilities obtained from  perturbation theory described in Section \ref{Sec:Smat}. Explicitly, for the muon to electron-neutrino oscillation case, each 
order of the  oscillation probability  is given by:
\begin{equation}
P^{(0)} (\nu_{\mu}\to \nu_{e})= 0,
\label{eq:result0}
\end{equation}
and,
\begin{equation}
P^{(1)} (\nu_{\mu}\to \nu_{e})= 4\frac{ \left|\Sigma\right| ^2 s_{23}^2 }{r_A^2 \eta^2}\sin^{2} \left(\frac{\Delta_{31} x}{2} r_A \eta \right),
\label{eq:result1}
\end{equation}
and,
\begin{equation}
P^{(3/2)}(\nu_{\mu}\to \nu_{e})=\frac{8 c_{23} s_{23} |\Sigma||\Omega|  \sin \left(\frac{\Delta_{31} x}{2}     r_A\Gamma\right) \sin \left(\frac{\Delta_{31}   x}{2}  r_A\eta\right) \cos \left(\frac{\Delta_{31} x }{2} 
r_A\Lambda -\phi _{\Sigma }+\phi _{\Omega }\right)}{ r_A^2\Gamma  \eta},
\label{eq:result32}
\end{equation}
and,
{\small
\begin{eqnarray}
P^{(2)} (\nu_{\mu}\to \nu_{e})&=&\frac{4c_{23}^2|\Omega|^2\sin^2\left(\frac{\Delta_{31} x}{2}r_A\Gamma\right)}{(r_A\Gamma)^2}
\\&+&2  \left|\Sigma\right| ^2 s_{23}^2 \left(\frac{2 \left|\Sigma\right| ^2}{r_A^3\eta^3 }-\frac{r_\Delta s_{12}^2+2 s_{13}^2}{r_A^2 \eta^2}\right)(\Delta_{31} x)\sin \left(r_A\eta\Delta_{31} x \right)\nonumber\\
&-& 4  s_{23}^2  \left(\frac{4 \left|\Sigma\right| ^4}{r_A^4 \eta^4}-\frac{2 \left|\Sigma\right|^2  \left(r_\Delta s_{12}^2+2 s_{13}^2\right)}{r_A^3 \eta^3}
+\frac{\left|\Sigma\right| s_{13} \left(2 r_\Delta  s_{12}^2+s_{13}^2\right) \cos \left(\delta_{\rm CP}+\phi _{\Sigma }\right)}{r_A^2 \eta^2}\right)\nonumber \\
&\times&\sin ^2\left(\frac{\Delta_{31}\,  x r_A  \eta}{2}  \right) +4 c_{23} \left|\widetilde{\epsilon}_{\mu \tau }\right|\left|\Sigma\right| ^2 s_{23} \sin \left(\frac{\Delta_{31} x}{2}   r_A  \eta\right)
\nonumber\\
&\times&
\left(\frac{ \sin \left(\widetilde{\phi}_{\mu \tau }
-\frac{\Delta_{31} x\, r_A  (\Gamma +\Lambda )}{2}   \right)}{r_A^2   \eta \Gamma  \Lambda}
 -\frac{\sin \left(\widetilde{\phi}_{\mu \tau }-\frac{\Delta_{31} x\,  r_A  \eta}{2}  \right)}{ r_A^2 \eta^2 \Gamma}+\frac{\sin \left(\tilde{\phi}_{\mu \tau }
 +\frac{\Delta_{31} x\,   r_A  \eta}{2} \right)}{  r_A^2 \eta^2\Lambda}\right)\nonumber , 
\label{eq:result2}
\end{eqnarray}
}
where we have defined 
\begin{eqnarray}
\Sigma&=&|\Sigma|e^{i\phi_{\Sigma}}\equiv s_{13}e^{-i\delta_{\rm CP}}+r_A{\widetilde{\epsilon}_{e\tau}}, \nonumber \\
\Omega&=&|\Omega| e^{i\phi_\Omega}\equiv r_\Delta c_{12}s_{12}+r_A\widetilde{\epsilon}_{e\mu}, \nonumber\\ 
 \Lambda&\equiv&\frac{1}{r_A}+\widetilde{\epsilon}_{\tau\tau},
\nonumber\\
\Gamma&\equiv&(1+\widetilde{\epsilon}_{ee}), \nonumber\\
\eta&\equiv&\Lambda -\Gamma.
\label{Eq:lambdas}
\end{eqnarray}

Our {\it  perturbative} probability it will be the sum of Eqs.~(\ref{eq:result0}-\ref{eq:result2}),  where Eq.~(\ref{Eq:lambdas}) is implicitly assumed, which results in:
\begin{eqnarray}
P^{\rm perturbative} (\nu_\mu\to \nu_e)\equiv  &=&~P^{(0)} (\nu_{\mu}\to \nu_{e})
+P^{(1)} (\nu_{\mu}\to\nu_{e})\nonumber\\
&+&P^{(3/2)} (\nu_{\mu}\to \nu_{e})+P^{(2)} (\nu_{\mu}\to \nu_{e}).
\label{Eq:Pperturb}
\end{eqnarray}
Our final result,  Eq.~(\ref{Eq:Pperturb}), it is the perturbative solution for the full neutrino evolution equation and  it is a power-law series of our expansion parameters, $(s_{13}, r_{\Delta},\widetilde{\epsilon}_{e\mu},\widetilde{\epsilon}_{e\tau},\widetilde{\epsilon}_{\mu\tau})$. 
Due to our choice of NSI hierarchy, we can rewrite as the combination
$(\Sigma, \Omega,r_{\Delta},\widetilde{\epsilon}_{\mu\tau})$. We can related $\Sigma$ and $\Omega$ with entries of perturbated Hamiltonian, $\widetilde{H}^{ (a)}$ and $\widetilde{H}^{(b)}$.
To obtain the anti-neutrino oscillation probabilities, in Eq.~(\ref{Eq:Schr01})  we change  $U\rightarrow U^{*}$,  $r_{A}\rightarrow -r_{A}$ and $\epsilon \rightarrow \epsilon^{*}$, whose implies in the  modifications in the effective parameters given  as
\begin{eqnarray}
\Gamma &\rightarrow& \Gamma, \nonumber \\
\Sigma &\rightarrow& \overline{\Sigma} = s_{13}e^{i\delta_{\rm CP}}-r_A{(\widetilde{\epsilon}_{e\tau})^{*}}, \nonumber \\
\Omega &\rightarrow& \overline{\Omega} = r_\Delta c_{12}s_{12}-r_A(\widetilde{\epsilon}_{e\mu})^{*}.
\label{Eq:anti_lambdas}
\end{eqnarray}
The formulas given by Eqs.~(\ref{eq:result1}-\ref{eq:result2}) can be directly applied to any long-baseline experiment. 

\begin{figure}[ht!]
\hskip -1.cm
\includegraphics[scale=0.7]{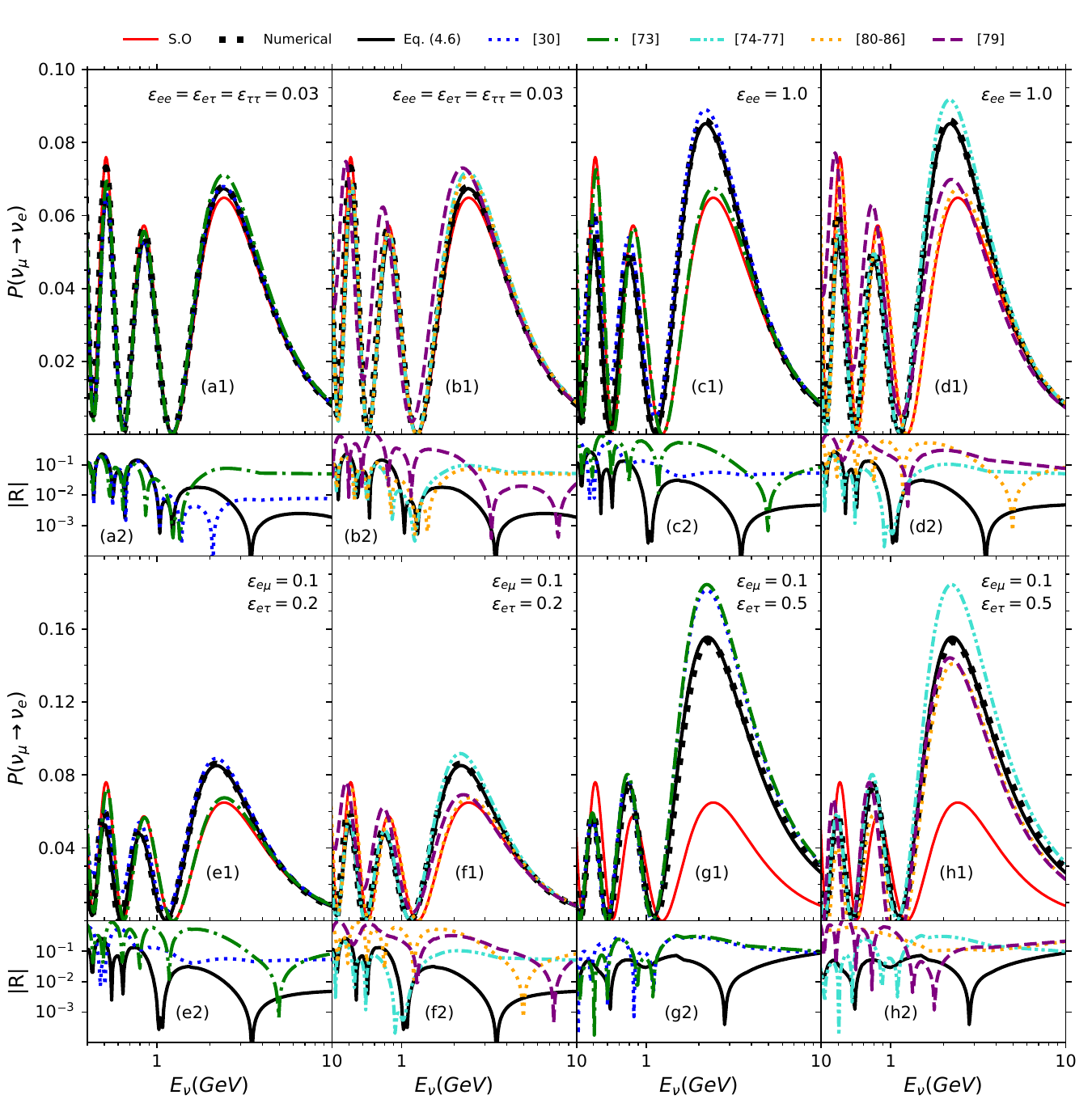}
\vskip -0.5cm
\caption{{\bf Panels  (a1), (b1), (c1), (d1), (e1), (f1), (g1), h1):} Comparison between the numerical solution and the results from perturbation theory for neutrino  oscillation probability. Our perturbative solution (solid black curve) is obtained using Eq.~(\ref{Eq:Pperturb}) and our numerical results are represented by the big dotted black curve. Also, the standard oscillation solution is shown by the solid cyan curve. Other colored dashed lines refers to the predictions from the Refs. quoted in Table~\eqref{tab:dist}. {\bf Panels  (a2), (b2), (c2), (d2), (e2), (f2), (g2), (h2):} The ratio of probabilities  as it is given in Eq.~\eqref{Eq:ratio} for the respective NSI values assumed in the panels (a1), (b1), (c1) and (d1). Notice the change in the scale of the plots in the upper/lower panel. We assume the values of mixing angles given in the text and L=$1300$km. All NSI parameters are assumed to be real.}
\label{fig:mina}
\end{figure}

In Figure~\ref{fig:mina} we made the comparison of $P(\nu_{\mu}\to \nu_{e})$  muon neutrino to electron neutrino for our results,shown in black line,  given by Eq.~(\ref{Eq:Pperturb}) and compare with the results of other perturbative solutions present in the literature~\cite{Asano:2011nj,kikuchi2009perturbation,Meloni:2009cg,Kopp:2007ne,Blennow:2016etl,Liao:2016hsa,Deepthi:2016erc,Deepthi:2017gxg,Dey:2018yht,Ge:2016dlx,Meloni:2009ia,Chatterjee:2014gxa,Masud:2015xva,Masud:2016bvp,Masud:2016gcl}. We also compare it  with the full numerical solution of neutrino evolution, shown in big dotted black curve,  given in Eq.~(\ref{Eq:Schr02a}). For this figure we assume $L=1300$ km  and  the matter density is assumed to be $\rho=2.8$ g/cm$^{3}$. We show in panel (a1), (c1), (e1) and (g1) the class of models of Ref.~\cite{Asano:2011nj,kikuchi2009perturbation} and in the panels (b1), (d1). (f1) ad (h1) the class of models~\cite{Meloni:2009cg,Kopp:2007ne,Blennow:2016etl,Liao:2016hsa,Deepthi:2016erc,Deepthi:2017gxg,Dey:2018yht,Ge:2016dlx,Meloni:2009ia,Chatterjee:2014gxa,Masud:2015xva,Masud:2016bvp,Masud:2016gcl} with different color lines as shown in
Figure~\ref{fig:mina}.

The relative difference between the full numerical solution and the different analytical models are defined by
\begin{equation}
R=\frac{P^{\rm model}-P^{\rm numerical}}{\overline{P}^{\rm numerical}},
\label{Eq:ratio}    
\end{equation}
where $P^{\rm model}$ is the oscillation probability for different perturbation theory scenarios presented in  Table~\ref{tab:dist}, $P^{\rm numerical}$ is the full numerical computation and $\overline{P}$ represents the mean probability in nearest neutrino bin energies\footnote{Strictly speaking, we applied the median filter from  Python~\cite{scypy}. 
The averaging procedure is used to avoid divergences in the ratio of $ R $ when the numerical probability has a local minimum.
}.  We showed in Figure~\ref{fig:mina}, the ratio R in the panel (a2) that correspond to the models shown in the panel (a1). The same applies for the other panels  (b2), (c2), (d2), (e2), (f2), (g2), and (h2).  Let us first comment on panel (a2). All curves agree with the numerical computation for energies above 1 GeV. The region below 1 GeV is where it happens the stronger effects of the  {\it solar scale} for long-baseline experiments, and this disagreement is known already.  Nowadays, we know that this is related to the standard oscillation scenario, and here we will refer only to the results for energies above 1 GeV.  For the other panels, we choose values of NSI parameters that can exemplify better the range of applicability of different perturbative solutions of the neutrino probability. If we follow our results, shown by the black curve in all panels showing the ratio R, we can notice it have the better agreement with respect with all other curves for any of the set of the NSI parameters chosen.  In the intermediate panels  (a2) and (b2), the ratio R corresponds to what we define as {\it small} NSI parameters; our results are equal or better than the other perturbative series. When we increase the diagonal NSI element, which differentiates the models as well in the Table~\ref{tab:dist}, the panel (c2) and (d2) show that our results still can reproduce the numerical results, smaller R, and the others begin to deviate more badly from the numerical solution reaching a  $R\sim 20\% $. For the panels (e2) and (f2), we have shown the results for larger values of non-diagonal NSI parameters, where we assume to be real, $\epsilon_{e\mu}=0.1$ and $\epsilon_{e\tau}=0.2$, but still, our computation have a much better agreement.  For even larger values of,  where we assume to real, 
 $\epsilon_{e\mu}=0.1$ and $\epsilon_{e\tau}=0.5$, shown in the panels (g2) and (h2) we can begin to notice in there are models with $R\sim 30\%$. 
The summary is that for any of these choices of NSI parameters, we can have a better agreement than any other model with the numerical results of the neutrino probability.

The same approach within perturbation theory can be applied to calculate muon neutrino survival probability. In 
Appendix~\ref{Ap:Pmm} we show our results for muon neutrino probability
$P^{\rm perturbative}(\nu_{\mu}\to\nu_{\mu})$  in the perturbative approach.

\section{Study of  degeneracies in the neutrino probability }
\label{Sec:degenerate}

 As mentioned in Section \ref{Sec:mot}, it was found that in both standard scenarios as well as in the NSI scenario that we can have multiple solutions of the parameters that led to the same the neutrino oscillation probability.  Examples of this are in  Ref.~\cite{Coloma:2016}, where the neutrino oscillation probability calculated numerically and that it will be relevant later.  Another example is when in  Ref.~\cite{kikuchi2009perturbation}, it was analyzed the conditions for {\it matter hesitation} when the NSI neutrino probability has the same value as the vacuum standard oscillation probability.

Here, we choose to use the advantage of having  an analytical expression for neutrino probability and use this to understand the root of the neutrino probability degeneracy. The degeneracy studied in the literature can be classified in the three categories: 

\begin{enumerate}
\item when it involve only  standard oscillation parameters,
\begin{equation}
\left[ P (\nu_{\alpha}\to \nu_{\beta})({\pmb{\Delta m_{ij}^2}},{\pmb{\theta_{ij}}})\right]^{(\rm SO))}=
\left[ P(\nu_{\alpha}\to \nu_{\beta})({\Delta m_{ij}^2}, ,\theta_{ij})\right]^{(\rm SO)} ,
\label{Eq:conddeg1a}
\end{equation}
where $\left[P (\nu_{\alpha}\to \nu_{\beta})({\pmb{\Delta m_{ij}^2}},{\pmb{\theta_{ij}}})\right]^{(\rm SO))}$ is the three-neutrino standard neutrino probability for a given set of mixing angle parameters $({\pmb{\Delta m_{ij}^2}},{\pmb{\theta_{ij}}})$ and $({\Delta m_{ij}^2},\theta_{ij})$ are different set of parameters. One example where this equation it is fulfilled is a special case of  the generalized mass ordering degeneracy for null NSI parameters~\cite{Coloma:2016}, as mentioned in Section \ref{Sec:intro}. It is the so called matter hesitation phenomena, where the neutrino probability in matter have the same dependence with energy as it was in vacuum.  It is an example of this equality expressed in Eq.~(\ref{Eq:conddeg1a}) and it was first discussed in 
Ref.~\cite{kikuchi2009perturbation}. 
  We will not consider the  degenerescence type I, as they have been discussed elsewhere.
 
\item a second case it is when it involves only NSI parameters, 
\begin{equation}
\left[ P (\nu_{\alpha}\to \nu_{\beta})(\widetilde{\epsilon}_{\alpha\beta},\Delta m_{ij}^2,\theta_{ij})\right]^{(\rm NSI))}=
\left[ P (\nu_{\alpha}\to \nu_{\beta})(\pmb{\widetilde{\epsilon}_{\alpha\beta}},{\Delta m_{ij}^2},\theta_{ij})\right]^{(\rm NSI)} ,
\label{Eq:conddeg1b}
\end{equation}
where $\left[ P (\nu_{\alpha}\to \nu_{\beta})(\widetilde{\epsilon}_{\alpha\beta},\Delta m_{ij}^2,\theta_{ij})\right]^{(\rm NSI))}$ is the neutrino probability in the NSI scenario, $\pmb{\widetilde{\epsilon}_{\alpha\beta}}$ and $\widetilde{\epsilon}_{\alpha\beta}$ are two different set of NSI parameters.
One example of this behaviour was shown in  Ref.~\cite{Friedland:2004ah}. This is the more oversight case in the literature and it can have impact on the understand of the future results of neutrino oscillations. We will discuss later in detail 
in Section \ref{degeii} and in the respective Figures~\ref{fig:NSI-DUNE}, \ref{fig:DEG-P}, and \ref{fig:DEG-regi},  the source of this behaviour using our perturbative formulas.

\item   when is due to the standard oscillation parameters and NSI parameters.
Respectively 
\begin{equation}
\hspace{-1.1cm}\left[ P (\nu_{\alpha}\to \nu_{\beta})(\pmb{\widetilde{\epsilon}_{\alpha\beta}} ,\Delta m_{ij}^2,\theta_{21},\theta_{13},\pmb{\theta_{23}})\right]^{(\rm NSI)}
=
\left[ P (\nu_{\alpha}\to \nu_{\beta}) (\widetilde{\epsilon}_{\alpha\beta} = 0 ,\Delta m_{ij}^2,\theta_{21},\theta_{13},\theta_{23})\right]^{(\rm SO)} ,
\label{Eq:conddeg1c}
\end{equation}
where 
$\left[ P 
(\nu_{\alpha}\to \nu_{\beta})(
\pmb{\widetilde{\epsilon}_{\alpha\beta}},\Delta m_{ij}^2,\theta_{21},\theta_{13},\pmb{\theta_{23}})\right]^{(\rm NSI))}$ is the 
neutrino oscillation probability in the NSI scenario,
$\left[ P (\nu_{\alpha}\to \nu_{\beta}) (\widetilde{\epsilon}_{\alpha\beta} = 0 ,\Delta m_{ij}^2,\theta_{21},\theta_{13},\theta_{23})\right]^{(\rm SO)}$ is the neutrino oscillation probability in the standard scenario. Here $\pmb{\theta_{ij}}$  and $\theta_{ij}$ are two different values for each mixing angle i,j=1,2,3. 
This example of  degenerescence was discussed in many situations, one was the so called {\it dark solution for the solar neutrino problem}~\cite{Miranda:2004nb}.  It was generalized for any oscillation scenario in  Ref.~\cite{Coloma:2016}. We will discuss it in  Section \ref{eqqtudo} and 
Figure~\ref{fig:degeneracycomp}. 
\end{enumerate}

Furthermore, in the NSI scenario, the parameters which describe the neutrino oscillation are: 
\begin{enumerate}
    \item Standard Oscillation parameters:  six SO parameters (the three mixing angles $\theta_{12}$, $\theta_{13}$, $\theta_{23}$, the two squared mass differences, $\Delta m^{2}_{21}$ and $\Delta m^{2}_{31}$, and the CP phase, $\delta_{\rm CP}$),
    \item  Non-Standard Oscillation parameters: two real parameters,  $\epsilon_{ee}$, and $\epsilon_{\tau \tau}$ and  three complex parameters $\epsilon_{e\mu}$, $\epsilon_{e\tau}$, and $\epsilon_{\mu\tau}$,
\end{enumerate}
and then we have in total fourteen parameters.
When perturbation theory is applied to solve  Eq.~(\ref{Eq:Schr01}), we can describe the neutrino oscillation probability as function of 
\begin{enumerate}
\item the parameters, $(\Lambda, ~\Gamma,~\Sigma,~\Omega)$  given in Eq.~(\ref{Eq:lambdas}),
\item the parameter $\epsilon_{\mu\tau}=|\epsilon_{\mu\tau}|e^{i\phi_{\mu\tau}}$,
\item the parameters $s_{13}$, $s_{12}$, $s_{23}$ and,
$\Delta_{31}$ and, $r_{\Delta}$,
\end{enumerate}
and then we have  eleven free parameters in this framework.

There are in the literature distinct theoretical works where the degeneracy problem was studied. Noticeable examples are 
in  Refs.~\cite{Liao:2016hsa,Yasuda:2007jp,Ge:2016dlx, Masud:2015xva, Masud:2016bvp}, where the degeneracy between NSI parameters and different mixing parameters are studied. Indeed, in  Ref.~\cite{Liao:2016hsa} it is assumed the  real parameter $\epsilon_{ee}\in H^{(0)}$ and the interplay between $\delta_{\rm CP}$ and $\epsilon_{e\mu}$ and  the  magnitude of complex parameter $|\epsilon_{e\tau}|$ is considered in the calculation of DUNE sensitivities. 
Our Eqs.~(\ref{Eq:gamma}, \ref{Eq:a}, \ref{Eq:NSIH12A}) present  a similar relation between the above mentioned NSI parameters and the standard oscillation parameters, as it can be verified in the Eqs.~($C1$, $C2$) of  \cite{Yasuda:2007jp} and the associated text. In Ref.~\cite{Masud:2015xva}, the resulting oscillation formula presented in their  Eq.~(12) depends on explicitly  combinations of $\delta_{\rm CP}$ and the phases associated with the   parameters that are considered  with order of the magnitude of NSI parameters $|\epsilon_{e\mu}|, |\epsilon_{e\tau}| \le 0.1 \approx \sqrt{\kappa} $. All the other NSI parameters are assumed to the order $\kappa$ and the impact on the DUNE sensitivities to  the $\delta_{\rm CP}$ is calculated. 
In  Ref.~\cite{Masud:2016bvp} the authors assume the neutrino oscillation probabilities  given in  Refs.~\cite{Kimura:2002hb,Kimura:2002wd, Ohlsson:2013ip}  and calculate the DUNE sensitivities in terms that are  odd and even  terms with respect to $\delta_{\rm CP}$. In  Ref.~\cite{Masud:2016gcl} a similar procedure is used to determine DUNE sensitivities to the neutrino mass hierarchy.   Additionally, combinations of  $\delta_{\rm CP}$ + NSI phases also appear in the results in  Ref.~\cite{Ge:2016dlx}.

Moreover, it the sensitivity studies in the plane of the NSI parameters; it appears as multiple solutions in numerical computations.
One example, in  Ref.~\cite{Gouvea:2015}, in the plane $|\epsilon_{e\tau}|\times \epsilon_{ee}$ there are two branches of solutions, {\it butterfly} like and also $\epsilon_{\tau\tau} \times \epsilon_{\mu\tau}$ plane. Similarly, multiple solutions found in  Ref.~\cite{Coloma:2016}.
 One set of solutions it compatible with {\it small} NSI parameters, but also it has a puzzling behavior {\it large} NSI parameters that also are compatible. In these sensitivity studies, they compute the rate of events in the case of standard oscillation probability and compare with the NSI probability. Due to this, we were motivated to ask
if the perturbation theory can explain (or at least mimic) the existence of these multiple solutions.

To answer such question, in what follows, we will use oscillation formulas within perturbation theory in two different ways:
\begin{itemize}
\item Perform a graphic comparison between perturbative  solutions between different NSI assumptions,
\item  Find analytic expressions for the neutrino probability degenerescence from Eqs.~(\ref{Eq:lambdas}-\ref{eq:result2}).
\end{itemize}

\subsection{Degenerescence type 2, only NSI parameters}
\label{degeii}

\begin{figure}[ht!]
\hskip -1cm
\includegraphics[scale= 0.78]{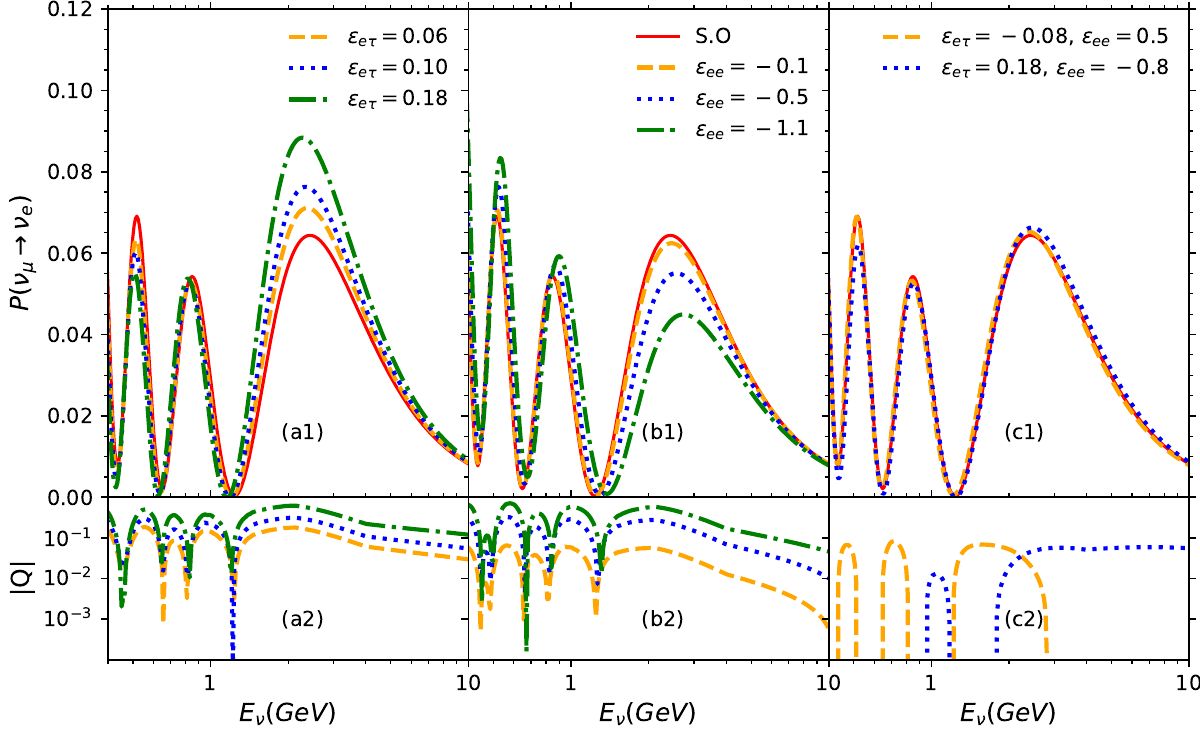}
\caption{
{\bf Panels (a1), (b1), (c1):} Comparison between our analytical solution for  the oscillation probabilities, $P(\nu_{\mu} \to \nu_e)$, for the SO (solid line) case  and different sets of  $\epsilon_{ee}$ and $\epsilon_{e\tau}$ NSI parameters (doted and doted-dashed lines). In panel (a1) we show the SO prediction and three NSI cases where only $\epsilon_{e\tau}$ is non-zero and assumes the values indicated in the plot. In panel (b1) we show SO prediction and three cases where only $\epsilon_{ee}$ is non-zero. In panel (c1) we show SO prediction and two cases where only $\epsilon_{ee}$ and $\epsilon_{e\tau}$ are non-zero and with opposite sign.
{\bf Panels (a2), (b2), (c2):} We shown the ratio $Q$ as it is given in Eq.~\eqref{Eq:Qratio} for the correspondent curves quoted in panels (a1), (b1), (c1).  We assume in this plot that all NSI parameters are real.
}
\label{fig:NSI-DUNE}
\end{figure}

As a matter of illustration of such degenerate behavior of neutrino time evolution Hamiltonian, in the panels of Figure~\ref{fig:NSI-DUNE} where we show the  effect of the diagonal NSI $\epsilon_{ee}$ parameter and the non-diagonal NSI parameter, which we assume to be real for this figure, $\epsilon_{e\tau}$.  In panels (a1), (b1) and (c1), we show the muon neutrino to electron neutrino conversion. In panels (a2), (b2) and (c2) we graph the quantity
\begin{equation}
    Q=\frac{P^{\rm NSI}-P^{\rm SO}}{\overline{P}^{\rm SO}},
\label{Eq:Qratio}
\end{equation}
that was calculated in the same way as in Eq.~\eqref{Eq:ratio}. Nevertheless, now, we have $P^{\rm NSI}$ that refers to  Eq.~\eqref{Eq:Pperturb} including the respective non-standard interaction, and $P^{\rm SO}$ is the same equation in the standard oscillation scenario.

We noticed in the left (middle) panel of Figure~\ref{fig:NSI-DUNE} that due to, assumed real parameter, $\epsilon_{e\tau}>0$ ($\epsilon_{ee}<0$)  has the effect to  decrease (increase) of the oscillation amplitude. For this set of parameters, when the only non-zero parameters are, assumed here to be real,  $\epsilon_{e\tau}$ and $\epsilon_{ee}$, we have
\begin{equation}
\widetilde{\epsilon}_{\mu\mu}=\widetilde{\epsilon}_{\tau\tau}=\widetilde{\epsilon}_{\mu\tau}=0;\quad
\widetilde{\epsilon}_{e\mu},\widetilde{\epsilon}_{e\tau},\widetilde{\epsilon}_{ee}\neq 0.
\label{Eq:Example_NSI}
\end{equation}
From Eq.~(\ref{Eq:lambdas}) and  Eq.~(\ref{Eq:Example_NSI})  implies that 
\begin{equation*}
\Omega\to \frac{1}{r_A}; \quad \Gamma\to 1+\epsilon_{ee}; \quad\eta=\frac{1}{r_A}-(1+\epsilon_{ee});\quad\Sigma \sim r_A c_{23}\epsilon_{e\tau};\quad \Omega\sim - r_A s_{23}\epsilon_{e\tau}.   
\label{siga}
\end{equation*}
Then if, assumed to be here a real parameter,  $\epsilon_{e\tau}>0  (\epsilon_{ee}<0)$ implies a larger $\Sigma (\eta)$. From  the neutrino probability in the first order, Eq.~\eqref{eq:result32},  larger $\Sigma (\eta)$ implies in a smaller (larger) amplitude of neutrino oscillation. For lower values of energy, the high orders of the perturbation theory are important and they depend also on $\Omega$ that have a different energy dependence from $\Sigma$.
Then even one single NSI parameter can have opposite effects in the oscillation probability and can indicate that the combined effect of more than one NSI can cancel the NSI effect.  To illustrate that, in the
right panel of Figure~\ref{fig:NSI-DUNE}  we show two cases where  $\epsilon_{ee}$ and $\epsilon_{e\tau}$ are intentionally large and with opposite signs. In this situation for both non-zero, it happens that $\Sigma$ and $\eta$ decrease (increase) when $\epsilon_{e\tau}$ and  $\epsilon_{ee}$ have opposite signs as you can see in Eq.~\eqref{siga}, and for a particular choice we can have 
\begin{equation}
|\Sigma| \sim |\eta|.
\end{equation}
Also,   small $\eta$ values do not change the phase of neutrino oscillation (that translate into the condition  $\epsilon_{ee}<1$). As one can see, in this situation we have the canceling the NSI effects in  Eq.~\eqref{eq:result32}. For lower energies, the contribution of high orders is more important and the effects of $\Omega$ are important and there is no canceling effects.
 In what follows, we apply the perturbation theory to determine analytically the origin of the neutrino oscillation probabilities degenerescence. 

\subsection{Degenerescence type 3, standard oscillation mixing parameters and  NSI parameters}
\label{eqqtudo}
Although it is relatively easy to obtain the numerical solutions of Eq.~(\ref{Eq:Schr01}),  it is not straightforward to establish a cause-effect relationship for the given NSI parameters and the resulting modifications in the behavior of the neutrino time evolution. Henceforth, our formalism within perturbation theory can be used to interpret the resulting neutrino oscillation probabilities. 
To get the information about the sources of  the degenerate solutions for the neutrino probability we are going to use the fact that the two-neutrino flavor oscillation formula, 
\begin{equation}
P^{2\nu}_{\rm conversion}= \sin^2 2\theta \times \sin^{2}\left(\frac{\Delta m^2 L}{4E_{\nu}}\right),
\end{equation}
and the  first order of the perturbative series $P^{(1)}(\nu_{\mu} \to \nu_e)$ can be cast into a similar format
\begin{equation}
P^{(1)}(\nu_{\mu} \to \nu_e) = A^{(1)}(\nu_{\mu} \to \nu_e) \times \sin^{2}(\Phi^{(1)}(\nu_{\mu} \to \nu_e)),
\end{equation} 
where $A^{(1)}(\nu_{\mu} \to \nu_e)$ is the oscillation amplitude and $\Phi^{(1)}(\nu_{\mu} \to \nu_e)$ is the phase.

First of all, let us assume that NSI parameters are zero,
$\widetilde{\epsilon}_{\alpha\beta} = 0$ in
Eq.~(\ref{eq:result1}),  then we have the standard oscillation formula in our approach,
\begin{equation}
\left[
P^{(1)}(\nu_{\mu}\to \nu_{e})\right]^{(\rm SO)}=\left[A^{(1)}(\nu_{\mu}\to \nu_{e})\right]^{(\rm SO)} \sin^{2} \left(\frac{\Delta_{31} x}{2} (1-r_A) \right), 
\label{eq:P1SO}
\end{equation}
 where 
\begin{equation}
\left[A^{(1)}(\nu_{\mu}\to \nu_{e})\right]^{(\rm SO)}
= \frac{ 4s^2 _{13}  s_{23}^2 }{(1-r_A)^2 },
\label{novanopedaco}
\end{equation}
that correspond to the first term of the expansion of the probability for the mixing parameter $\theta_{13}$.

When we switched on the  diagonal NSI parameters that enter in $\widetilde{H}^{(0)}$,  then Eq.~(\ref{eq:result1}) can be recast  as 
\begin{eqnarray}
\left[
P^{(1)}(\nu_{\mu}\to \nu_{e})\right]^{(\rm diagonal\, NSI)}
&=&
\left[A^{(1)}(\nu_{\mu}\to \nu_{e})\right]^{(\rm diagonal\, NSI)}
\nonumber \\
&\times&\sin^{2} \left[\Phi^{(1)}(\nu_{\mu}\to \nu_{e})\right]^{(\rm diagonal\, NSI)}, 
\label{eq:P1H0}
\end{eqnarray}
where
\begin{eqnarray}
\left[A^{(1)}(\nu_{\mu}\to \nu_{e})\right]^{(\rm diagonal\, NSI)}&=&
\frac{ 4s^2 _{13}  s_{23}^2 }{\left[1-r_A+r_{A}(\widetilde{\epsilon}_{\tau\tau}-\widetilde{\epsilon}_{ee})\right]^{2}
} ,\nonumber \\
\left[\Phi^{(1)}(\nu_{\mu}\to \nu_{e})\right]^{(\rm diagonal\, NSI)}&=& \frac{\Delta_{31} x}{2} \left[1-r_A+r_{A}(\widetilde{\epsilon}_{\tau\tau}-\widetilde{\epsilon}_{ee})\right].
\label{eq:P1H0a}
\end{eqnarray}

Comparing Eq.~(\ref{novanopedaco}) and  Eq.~(\ref{eq:P1H0a}), we have 
\begin{eqnarray}
\frac{\left[A^{(1)}(\nu_{\mu}\to \nu_{e})\right]^{(\rm SO)}}{\left[A^{(1)}(\nu_{\mu}\to \nu_{e})\right]^{(\rm diagonal\, NSI)}} = (1+\gamma)^{2}, \nonumber \\
\frac{\left[\Phi^{(1)}(\nu_{\mu}\to \nu_{e})\right]^{(\rm SO)}}{\left[\Phi^{(1)}(\nu_{\mu}\to \nu_{e})\right]^{(\rm diagonal\, NSI)}} =\frac{1}{(1+\gamma)}, 
\label{Eq:ASOH0}
\end{eqnarray}
where we define real parameter $\gamma$
\begin{eqnarray}
    \gamma  & &\equiv \frac{r_{A}\left(\widetilde{\epsilon}_{\tau\tau}-\widetilde{\epsilon}_{ee}\right)}{1-r_{A}}. \nonumber
\label{Eq:gamma}
\end{eqnarray}

Now, we are switched on all NSI parameters, and we arrive to 
\begin{eqnarray}
\frac{\left[A^{(1)}(\nu_{\mu}\to \nu_{e})\right]^{(\rm diagonal\, NSI)}}{\left[A^{(1)}(\nu_{\mu}\to \nu_{e})\right]^{(\rm NSI)}}& =& \frac{1}{1+a\left(a+2\cos(\zeta)\right)}, \nonumber \\
\frac{\left[\Phi^{(1)}(\nu_{\mu}\to \nu_{e})\right]^{(\rm diagonal\, NSI)}}{\left[\Phi^{(1)}(\nu_{\mu}\to \nu_{e})\right]^{(\rm NSI)}} &=&1,
\label{Eq:PhiSOH0}
\end{eqnarray}
where the parameter a is defined as,
\begin{equation}
a \equiv \frac{r_{A}}{s_{13}}|\widetilde{\epsilon}_{e\tau}|,
\label{Eq:a}
\end{equation}
and  we  define $\zeta$, 
\begin{equation}
 \zeta \equiv \delta_{\rm CP} + \widetilde{\phi}_{e\tau}=\phi_{\Sigma},   \label{Eq:zeta}
\end{equation}
where the angle $\zeta$, it is a combination of the CP phase, $\delta_{\rm CP}$, and the NSI phase from $\widetilde{\epsilon}_{e\tau}$. It is equal to the phase of $\Sigma$ parameter defined in Eq.~(\ref{Eq:lambdas}), $\Sigma=|\Sigma|e^{i\phi_{\Sigma}} $.
Notice that in Eq.~(\ref{Eq:PhiSOH0}) the modification in the oscillation amplitude due to the inclusion of NSI depends only the net phase $\zeta$,  and does not depend on each phase individually. Indeed, also the oscillation probabilities given by  Eqs.~(\ref{eq:result1} and \ref{eq:result32}) depend only
the phase $\zeta$. Similar dependence on the sum of $\delta_{\rm  CP}$ and NSI phases was found in the Ref.~\cite{Ge:2016dlx}. Furthermore, from 
Eqs.~(\ref{Eq:ASOH0}-\ref{Eq:PhiSOH0})  one can determine the impact of NSI until first order in perturbation theory simply as
\begin{eqnarray}
\left[A^{(1)}(\nu_{\mu}\to \nu_{e})\right]^{(\rm NSI)}
 &=& \frac{1+a\left(a+2\cos(\zeta)\right)}{(1+\gamma)^{2}} \left[A^{(1)}(\nu_{\mu}\to \nu_{e})\right]^{(\rm SO)}~,
\label{Eq:NSIH12A}
\end{eqnarray}
and 
\begin{equation}
\left[\Phi^{(1)}(\nu_{\mu}\to \nu_{e})\right]^{(\rm NSI)} = (1+ \gamma)\left[\Phi^{(1)}(\nu_{\mu}\to \nu_{e})\right]^{(\rm SO)}~.
\label{Eq:NSIH12F}
\end{equation}
The Eqs.~(\ref{Eq:NSIH12A}) and (\ref{Eq:NSIH12F}) summarizes the modifications respectively in the terms of the amplitude $\left[A^{(1)}(\nu_{\mu}\to \nu_{e})\right]$ and the phase $\left[\Phi^{(1)}(\nu_{\mu}\to \nu_{e})\right]$ due to the inclusion of the NSI parameters that are present in Eq.~(\ref{eq:result1}). In what follow we investigate how our formalism within perturbation theory can be used to determine the origin of the degenerate behavior of neutrino time-evolution.


\begin{figure}[ht!]
\centering
\subfloat{\includegraphics[width=3.2in]{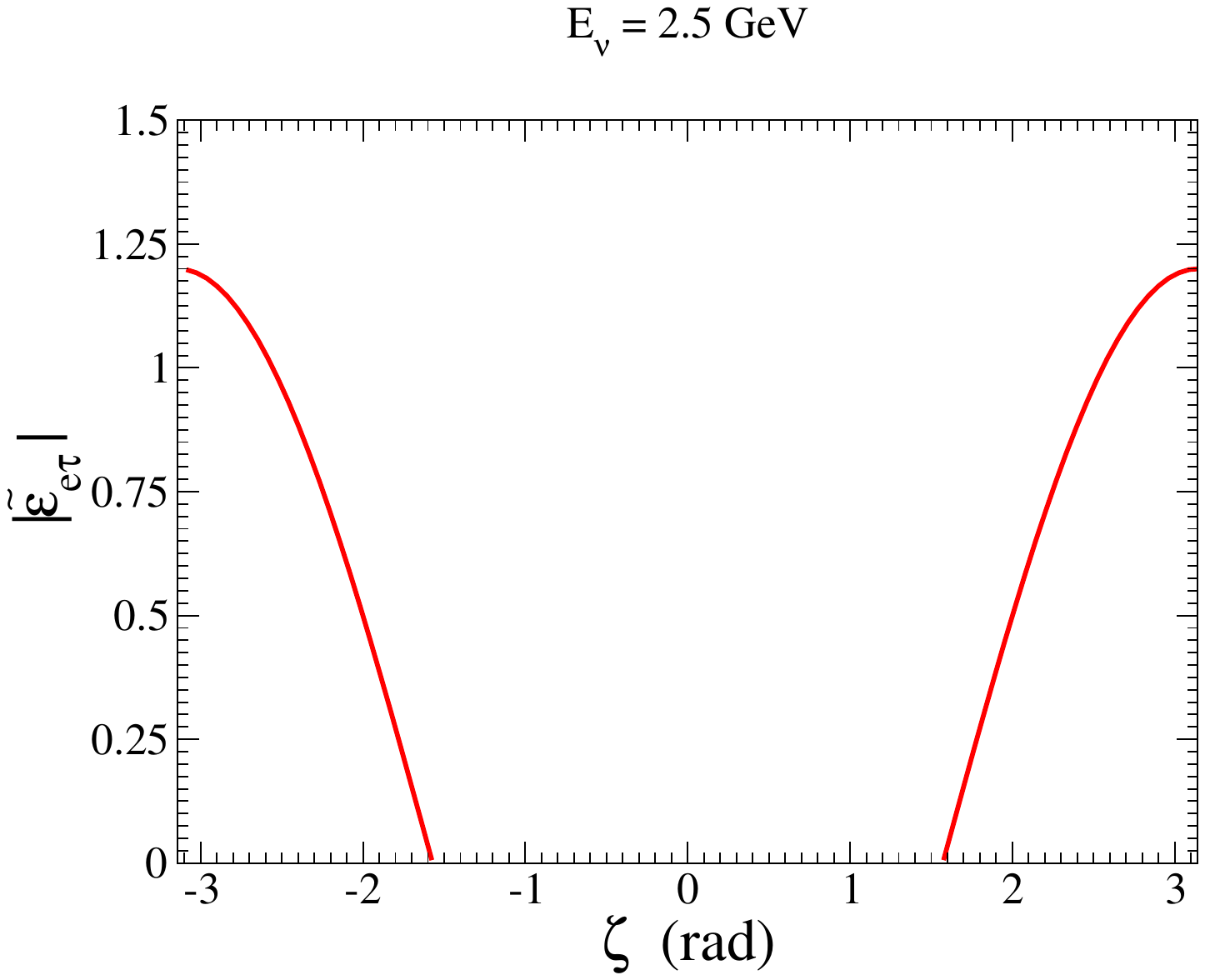}}\hskip-1.25cm
\subfloat{\includegraphics[width=3.2in]{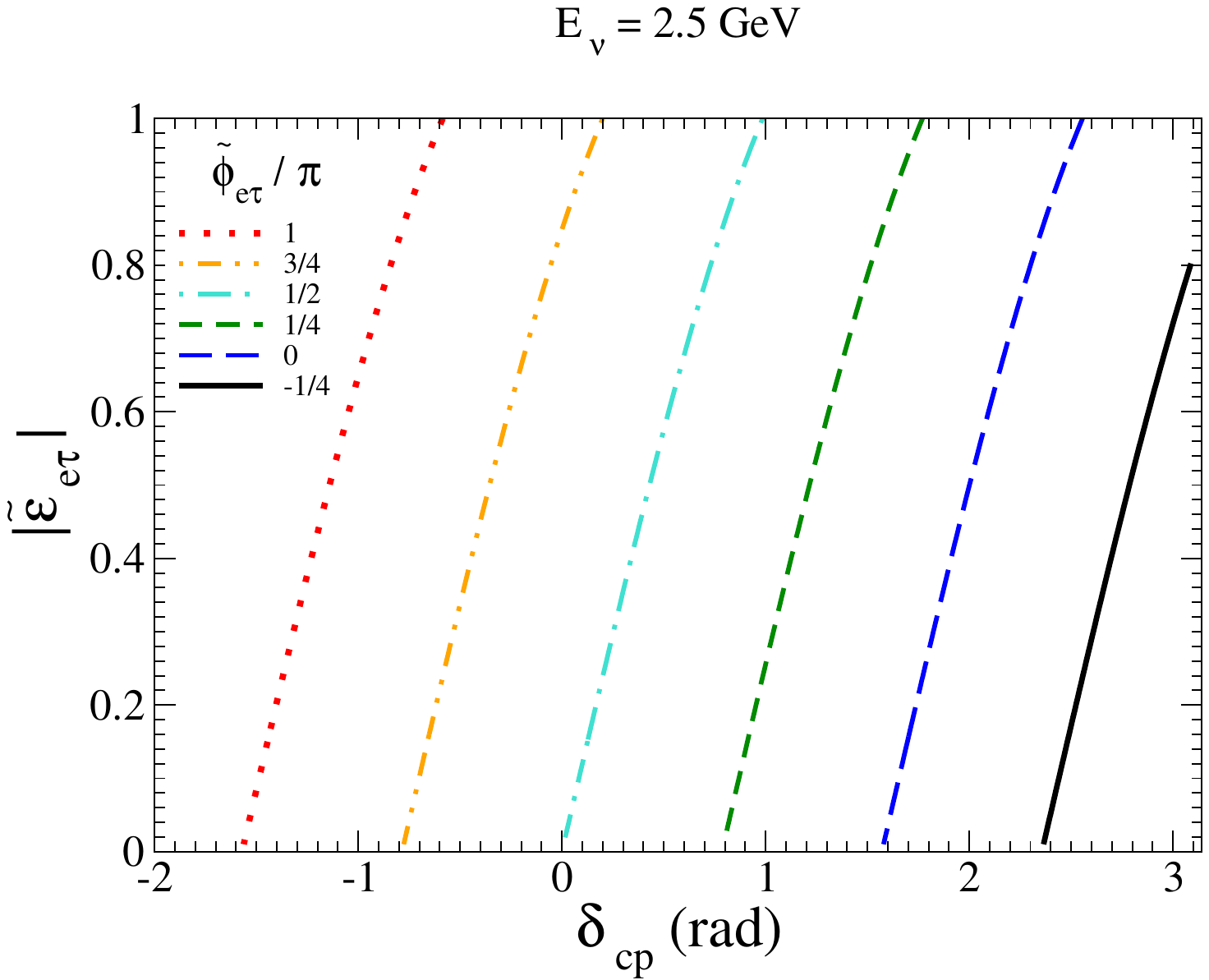}}
\caption{{\bf Left (right) panel:}, possible combinations  of $\zeta$ and $|\widetilde{\epsilon}_{e\tau}|$ ($\delta_{\rm CP}$ and  the amplitude of the NSI parameter  $|\widetilde{\epsilon}_{e\tau}|$) that 
lead to $
\left[
P^{(1)}(\nu_{\mu}\to \nu_{e})\right]^{(\rm SO)}=
\left[
P^{(1)}(\nu_{\mu}\to \nu_{e})\right]^{(\rm NSI)}$.  We assume all NSI parameters to be real. From the left-to-right we have decreasing values of phase $\tilde\phi_{e\tau}$  of the NSI element $\widetilde{\epsilon}_{e\tau}$.} 
\label{fig:DEG-P}
\end{figure}
The respective constraints over the standard oscillation and NSI parameters can be obtained from the Eq.~(\ref{Eq:NSIH12A}) and Eq.~(\ref{Eq:NSIH12F}). The conditions for both oscillation amplitude and phase to be degenerated implies in: 

\begin{enumerate}
    \item equal oscillation amplitude: from 
    Eq.~(\ref{Eq:NSIH12A}), $\gamma\to 0$
    \item equal oscillation phase: from 
    Eq.~(\ref{Eq:NSIH12F}), $a+2\cos (\zeta)\to 0$,
\end{enumerate}
where the first and second condition implies that the following condition on the real diagonal NSI parameters (i) 
\begin{equation}
\widetilde{\epsilon}_{\tau\tau}= \widetilde{\epsilon}_{ee},
\label{Eq:Eeta}
\end{equation}
and (ii)
\begin{equation}
|\widetilde{\epsilon}_{e\tau}| = -2\frac{s_{13}}{r_{A}}\cos{\zeta}.
\label{Eq:Eet}
\end{equation}

Both conditions warrant that the phase and the amplitude are equal for the SO oscillation and the SO $\otimes$  NSI cases,  for a given energy and when the first term of perturbative expansion is dominant.  
In the left (right) panel of Figure~\ref{fig:DEG-P} we
present the values in the plane 
$\widetilde{\epsilon}_{e\tau} \times \zeta $ ($\widetilde{\epsilon}_{e\tau} \times \delta_{\rm CP})$. We use  Eqs.~(\ref{Eq:NSIH12A}, \ref{Eq:NSIH12F}), where the conditions given in Eqs.~(\ref{Eq:Eeta}, ~\ref{Eq:Eet}) are obeyed, in such way that always we can find a solution for any $\widetilde{\epsilon}_{e\tau}$.  Notice that in the standard neutrino oscillation perturbation expansion, the first term of the expansion is independent of the CP violation, $\delta_{\rm CP}$, and here due to the NSI interference we are not sensitive to CP violation even at leading order.

Even more, from our results given in 
Figures~\ref{fig:mina} and \ref{fig:NSI-DUNE}, we can see that, for the DUNE case, the amplitude of neutrino oscillation tends to be more sensitive to the NSI than the oscillation phase. 
If we  impose the condition
\begin{eqnarray}
\left[A^{(1)}(\nu_{\mu}\to \nu_{e})\right]^{(\rm NSI)}
 &=& \left[A^{(1)}(\nu_{\mu}\to \nu_{e})\right]^{(\rm SO)}~,
\label{Eq:NSIH12V}
\end{eqnarray}
then this implies that from Eq.~(\ref{Eq:NSIH12A}) we define an effective phase $\zeta^{\prime}$ as
\begin{equation}
\cos \zeta^{\prime} = \frac{(1+\gamma)^2-1 - a^2}{2a},
\label{eqq}
\end{equation}
that it is written as function of $\gamma$ from  from Eq.~\eqref{Eq:gamma} and $a$ from Eq.\eqref{Eq:a}.
\begin{figure}[ht!]
\centering
\includegraphics[scale=0.55]{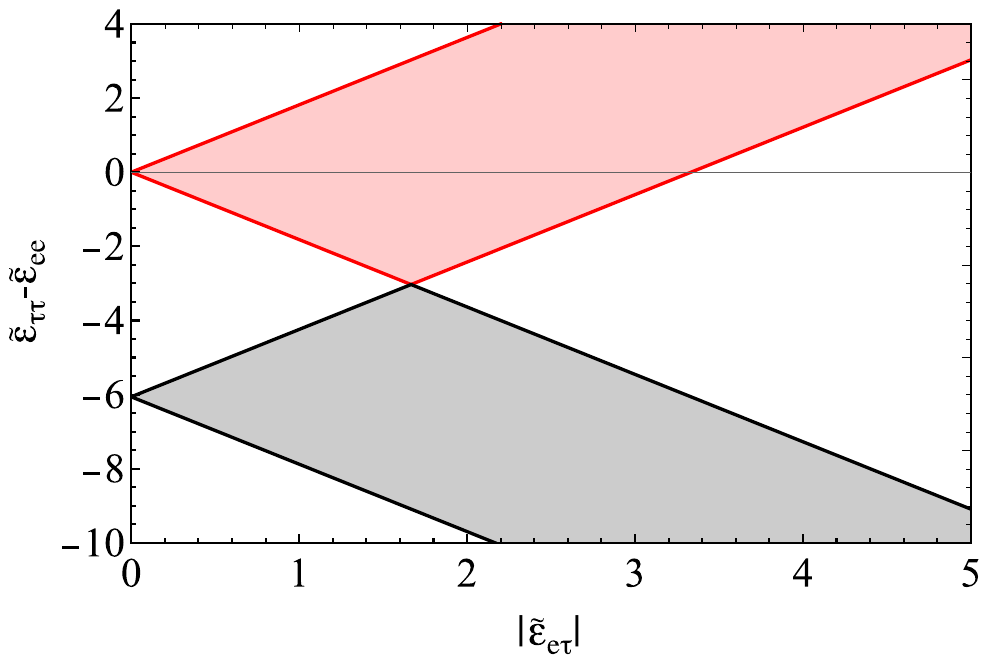}
\caption{Region in the phase-space of the magnitude of $|\widetilde{\epsilon}_{e\tau}|$, $\widetilde{\epsilon}_{ee}$ and $\widetilde{\epsilon}_{\tau\tau}$ that lead to have the same amplitudes 
$\left[A^{(1)}(\nu_{\mu}\to \nu_{e})\right]^{(\rm NSI)}=\left[A^{(1)}(\nu_{\mu}\to \nu_{e})\right]^{(\rm SO)}$. The non-diagonal NSI element, $\widetilde{\epsilon}_{e\tau}$ can be complex.}
\label{fig:DEG-regi}
\end{figure}
The  Eq.~\eqref{eqq} relate the variables, $\widetilde{\epsilon}_{e\tau},\widetilde{\epsilon}_{ee},\widetilde{\epsilon}_{\tau\tau}$ and $\zeta$.
By requiring that $-1 \le \cos(\zeta)\le +1$ we found the conditions  for the degeneracy in the amplitude happens if,
\begin{eqnarray}
|1-a|-1\le \gamma^{\prime}&\le& a, \nonumber \\
-(1+a)-1 \le \gamma^{\prime}& \le & -|1-a|-1 .
\label{Eq:deg-space}
\end{eqnarray}
The four conditions, due to the possibility to have $a>1$ or $a<1$, presented in Eq.~(\ref{Eq:deg-space}),  define a region of the phase-space where the amplitude of standard oscillation is degenerated with NSI in the plane of $\widetilde{\epsilon}_{e\tau},\widetilde{\epsilon}_{ee},\widetilde{\epsilon}_{\tau\tau}$. Then {\it we have found conditions to have the neutrino probability degeneracy between the standard neutrino scenario  and the  NSI scenario}.
The results  are presented in Figure~\ref{fig:DEG-regi}, where we can conclude that depending on the relative signal of 
$\widetilde{\epsilon}_{\tau\tau}-\widetilde{\epsilon}_{ee}$ and the values of $\widetilde{\epsilon}_{e\tau}$ we can have a complete canceling of the NSI effect. Then we have shown that we can have the cancellation of the NSI effect as exact for the first order of the perturbation theory. Higher-order effects of the perturbation theory can spoil the neutrino probability  degenerescence.

\subsubsection{Degenerescence type 3, assuming different mixing angles}
\label{deg:2}

\begin{figure}[ht!]
\centering
\includegraphics[width=12.0cm]{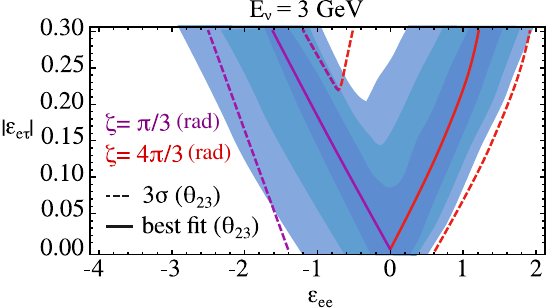}\\
\caption{Superposition of  our iso-probabilities (lines)  from Eq.~\eqref{Eq:Eet2} with the allowed region in the plane $\epsilon_{ee}$ and the amplitude $|\epsilon_{e\tau}|$ (shaded region) reported in Ref. ~\cite{Coloma:2016}. In this region,  color darkness refers to the $(1\sigma, 2\sigma, 3\sigma)$ regions  the authors found in their sensitivity study of the DUNE experiment. 
Solid (dashed) red  curves are generated using the best-fit point
for $\pmb{\theta_{23}}$, $\pmb{s_{23}^2}=0.441$ and for combined phase $\zeta=4\pi/3$ (rad) ( the  $3\sigma$ values $\pmb{s_{23}^2}=0.385\rightarrow 0.635$). The solid (dashed) magenta curves  
have the same respective $\pmb{\theta_{23}}$ values but the phase $\zeta$ is equal to $\zeta=\pi/3$ (rad). Both values for $\pmb{\theta_{23}}$ are  from  Ref.~\cite{esteban2017updated}.
}
\label{fig:degeneracycomp}
\end{figure}

Until now, we have assumed that the equality of probabilities for the same values of the mixing angles. For example,  Eqs.~\eqref{Eq:NSIH12A},  \eqref{Eq:NSIH12F}, and \eqref{Eq:NSIH12V}, and Figure~\ref{fig:NSI-DUNE}. 
Nevertheless, in the phenomenological analysis of the constraints on NSI parameters, we marginalized over the mixing parameters to show the constraints in the plane of NSI variables and then the degeneracy came from not from equality of probabilities in standard neutrino case and with NSI with  {\it the same} mixing angles.
One example of an analysis of the sensitivity of DUNE for NSI in the plane of $\epsilon_{e\tau}\times \epsilon_{ee}$  is  Ref.~\cite{Coloma:2016}. In this work~\cite{Coloma:2016}, it has marginalized overall mixing angles not shown in the plot. In this reference, one of the plots shown, reproduced in Figure~\ref{fig:degeneracycomp}, shows the blue region, going from darker blue, intermediate blue and lighter blue we have respectively  $1\sigma$,  $2\sigma$ and  $3\sigma$ allowed regions.     

We will try to understand this behaviour using our formalism of perturbation theory for NSI. We will assume that 
\begin{equation}
\left[ P (\nu_{\alpha}\to \nu_{\beta})(\pmb{\widetilde{\epsilon}_{\alpha\beta}} ,\pmb{\theta_{23}})\right]^{(\rm NSI))}
=
\left[ P (\nu_{\alpha}\to \nu_{\beta}) (\widetilde{\epsilon}_{\alpha\beta} = 0 ,\theta_{23})\right]^{(\rm SO)},
\label{Eq:conddeg}
\end{equation}
where for {\it right side of this equality} the standard oscillation case, we will fix the mixing angle $\theta_{23}$ at the best-fit point of  Ref.~\cite{esteban2017updated}, and for {\it the left side of this equality} we will assume a different angle $\pmb{\theta_{23}}$. We will search for these solutions of this equality, assuming a fixed $\theta_{23}$ and changing $\bm{\theta_{23}}$ inside the  $3\sigma$ allowed region \cite{esteban2017updated}.  A non-zero NSI will change the neutrino probability and we will search if we can use a different value of mixing angle  $\theta_{23}$, such that the value of NSI probability became equal to standard neutrino oscillation probability.

We can rewrite the Eq.~\eqref{Eq:conddeg}, using the explicit forms for the neutrino probability, as given in Eq.~\eqref{eq:result1}. After some algebra, discussed in detail in Appendix \ref{Ap:Deg} we arrive to the expression
\begin{eqnarray}
\frac{r^{2}_{A}}{s_{13}^2}|\widetilde{\epsilon}_{e\tau}|^{2}+2\frac{r_{A}}{s_{13}}|\widetilde{\epsilon}_{e\tau}|\cos(\zeta)+1
&=&\left(\frac{s_{23}}{\pmb{s_{23}}}\right)^2\left(\frac{B_1}{B_1+B_2r_A(\widetilde{\epsilon}_{\tau\tau}-\widetilde{\epsilon}_{ee})}\right)^2,
\label{Eq:Eet2}
\end{eqnarray}
where $\pmb{s_{23}}\equiv \pmb{\sin (\theta_{23})}$ is the value for $\theta_{23}$ for NSI neutrino probability in Eq.~\eqref{Eq:conddeg},  $\theta_{23}$ used in the standard neutrino probability and B1 and B2 are defined in the 
Appendix~\ref{Ap:Deg}.

The Eq.~(\ref{Eq:Eet2}) expresses the condition to have equality between the standard neutrino oscillation and the NSI probability
given in Eq.~(\ref{Eq:conddeg}) in terms of $\delta_{\rm CP}$, $\theta_{23}$ the complex parameter $|\widetilde{\epsilon}_{e\tau}|$, $\widetilde{\epsilon}_{ee}$, $\widetilde{\epsilon}_{\tau\tau}$ and the phase $\zeta=\tilde{\theta}_{e\tau}+\delta_{\rm CP}$. It also depends on the effective matter potential that neutrino crosses and the neutrino energy, which is contained in $r_{A}$ parameter. Concerning an example DUNE distance of  L=1300~km, in Figure~\ref{fig:degeneracycomp}, we show the iso-probability curves (solution of Eq.~  \eqref{Eq:Eet2}) for the parameters $\epsilon_{ee}$  and $\epsilon_{e\tau}$, assuming all other NSI parameters to be zero. We plot
\begin{enumerate}
    \item solving Eq.~\eqref{Eq:Eet2}, assuming values for $\pmb{\theta_{23}}$,  $\pmb{s_{23}}$ equal to best fit values. This implies that $\pmb{s_{23}}=s_{23}$. We assume that for combined phase  $\zeta=4\pi/3(\pi/3)$(rad) and we got the solid red (violet) curve respectively, 
    \item solving Eq.~\eqref{Eq:Eet2}, assuming values $\pmb{s_{23}}$ inside the $3\sigma$ allowed values of he   $\theta_{23}$. Again the solid red (violet) curve respectively for combined phase  $\zeta=4\pi/3(\pi/3)$(rad).
\end{enumerate}
The results of sensitivity for the NSI calculation for the DUNE experiment \cite{Coloma:2016}, are shown in Figure~\ref{fig:degeneracycomp}, where it was analyzed the sensitivity for the NSI, using the full three neutrino probability in matter, assuming that the test values of the standard neutrino oscillation and for the test the NSI probability. It was made the marginalization over the parameters $\Delta m^{2}_{31}$,
$\theta_{23}$, $\delta_{\rm CP}$ and the complex phase of $\phi_{e\tau}$. We superimposed these sensitivity regions (shaded regions)  in Figure~\ref{fig:degeneracycomp}.
From this comparison, we have that 

\begin{enumerate}
    \item (i) the central region at $1\sigma$ can describe as the region where the $\pmb{\theta_{23}} $ of NSI probability is equal to the $\theta_{23}$ of standard probability. The point $(\epsilon_{ee},\epsilon_{e\tau})=(0,0)$ correspond to the standard oscillation and for any value of  $\epsilon_{ee}$ and $\epsilon_{e\tau}$ non-zero correspond to NSI probability that it is degenerated with the standard probability. This canceling of $\epsilon_{ee}$ and $\epsilon_{e\tau}$ effects it is similar to what was discussed in 
    Section \ref{degeii},
    \item (ii) the external contours of the blue region can be understood as we can compensate for the large values of NSI by shifting the value of $\theta_{23}$ parameter, from the central value of the best-fit value to allowed at $3\sigma$ (shown by the dashed curves),
    \item the anti-correlation of $\epsilon_{ee}$ and $\epsilon_{e\tau}$ can be understood by the analytical discussion made in Section \ref{degeii} and also shown in Figure~\ref{fig:DEG-regi}. 
\end{enumerate}
We can conclude that we can reasonably understand the behavior of the multiple solutions for the NSI parameters, from the sensitivity analysis of the DUNE experiment in  Ref.~\cite{Coloma:2016}, as the freedom to change the mixing angle $\theta_{23}$ and the combined phase $\zeta$ in such way that we have the same numerical value of the NSI probability and standard probability even for large NSI diagonal  $\epsilon_{ee}$,

 \section{Conclusions}
\label{Sec:Conc}
We study the effects of inclusion of NSI in the perturbative approach to neutrino oscillations in the matter through the Dyson Series. We developed a perturbative approach where the diagonal NSI parameters are kept non-perturbative and the NSI non-diagonal parameters and the mixing parameters $\sin \theta_{13}$ and the ratio $r_{\Delta}$ as the perturbative parameters. We assume a hierarchy for NSI parameters, as $\widetilde{\epsilon}_{e\tau}$ as  sub-leading  parameter and $\widetilde{\epsilon}_{e\mu}$ and $\widetilde{\epsilon}_{\mu\tau}$ as   sub-sub-leading  parameters.  
This choice has the advantage that we can get all other cases of perturbative expansions in the literature as case limits of our case, as we can appreciate in Table~\ref{tab:dist}.  Using this approach we compute as an example, the neutrino oscillation as given in Eq.~(\ref{Eq:Pperturb}) for the muon neutrino to electron neutrino conversion probability, and in Eq.~(\ref{eq:Puuall}) for the muon neutrino survival probability.

In Figure~\ref{fig:mina}, we compare our analytical results (the black curve) with the numerical computation, shown in the dashed black line. The other results from perturbative approaches in the literature are also showed. For all choices of parameters, we are more trustworthy to the numerical computation than the other analytical perturbative computations available in the literature. Furthermore, as a by-product of our choice of NSI hierarchy is that the perturbated Hamiltonian in a block-diagonal format and this implies that we can describe the full probability as a function of   $(\Lambda, ~\Gamma,~\Sigma,~\Omega)$ functions defined in Eq.~(\ref{Eq:lambdas}).

Moreover, we have also have studied the underlying correlations between NSI parameters and mixing parameters that made there are multiple sets of NSI parameters and mixing angles that mostly have the same neutrino probability value. This degeneracy of the neutrino probability value that appears in numerical analysis in the literature, now it can be {\it explained} by the invariance of the amplitude and the phase of neutrino oscillation probability. This invariance was discussed in Section \ref{Sec:degenerate} for two cases,  (i) involving only NSI parameters and (ii) involving a combination of the NSI parameters and the standard oscillation parameters. We discussed the necessary conditions to have the neutrino probability degeneracy, and we apply for these two cases. In  Figure~\ref{fig:NSI-DUNE}, we have shown the reason when we have multiple NSI, we can have a canceling effect and effectively get the same probability as is the probability without NSI. In Figure~\ref{fig:degeneracycomp}, we show in the plane of $\epsilon_{ee}\times \epsilon_{e\tau}$ where we made a numerical analysis of the sensitivity of a given experiment for NSI parameter, and it appears these two branches of solutions. It follows from our discussion in 
Section \ref{Sec:degenerate}  the amplitude and the phase of $ \epsilon_{e\tau}$ are kept unchanged even we change the value of mixing angle $\theta_{23}$ and the NSI parameters. 

 Finally, the fact that now we have oscillation formulas that can apply to most of the  NSI allowed space of parameters should allow us to figure out the features of the future sets of experiments necessary to broken this degenerate behavior shown in the neutrino probabilities.  

\section*{Acknowledgments}

 O.L.G.P. is grateful for the support of FAPESP funding Grant  2014/19164-6, CNPq research fellowship 306565/2019-6 and 304715/2016-6, and  .  O.L.G.P. and  D.R.G. are grateful for partial support from the FAEPEX funding grant, No 2391/17. M. E. C. is grateful for the 140564/2018-7 and 130912/2016-6 funding from CNPQ. This study was financed in part by the Coordenação de Aperfeiçoamento de Pessoal de Nível Superior - Brasil (CAPES) - Finance Code 001.  The authors are thankful to  M. C. Gonzalez-Garcia and M. Maltoni, who kindly share the table of the $\Delta \chi^{2} \times \rm{NSI}$ parameters.

\bibliographystyle{JHEP}
\bibliography{NSI_arxiv_ii}

\appendix

\section{Explicit relation between NSI parameters in flavor basis and propagation basis}
\label{apa}
The rotation which relates flavor and propagation basis from Eq.~(\ref{Eq:basisH}) and with the redefinition given in Eq.~(\ref{mudanca})
implies in the following relation between NSI parameters:
\begin{eqnarray}
\widetilde{\epsilon}_{\mu\mu} & = &
s_{23}^2\epsilon_{\tau\tau}-2c_{23}s_{23} \Re(\epsilon_{\mu\tau})
\nonumber \\
\widetilde{\epsilon}_{\tau\tau} & = & 
c_{23}^2\epsilon_{\tau\tau} +2c_{23}s_{23} \Re (\epsilon_{\mu\tau})\nonumber \\
\widetilde{\epsilon}_{\mu\tau} & = & 
-s_{23}c_{23} \epsilon_{\tau\tau}+
(c_{23}^2-s_{23}^2)\Re(\epsilon_{\mu\tau}) +i \Im(\epsilon_{\mu\tau}) \nonumber\\ 
\widetilde{\epsilon}_{ee} & = &  \epsilon_{ee} 
\nonumber \\
\widetilde{\epsilon}_{e\mu} & = & c_{23}\epsilon_{e\mu} - s_{23}\epsilon_{e\tau} \nonumber \\
\widetilde{\epsilon}_{e\tau} & = & s_{23}\epsilon_{e\mu} + c_{23}\epsilon_{e\tau} 
\label{Eq:epslbasis}
\end{eqnarray}

\section{Formalism to perturbation theory for neutrino oscillations}
\label{Ap:Dyson}
The $S$ matrix  is responsible for the neutrino-state time evolution, 
\begin{equation}
|\nu_{\beta}(t)\rangle=S_{\beta\alpha}(t)|\nu_{\alpha}(0)\rangle,
\label{Eq:ketInt}
\end{equation}
where  the operator $S(t)$ is given by, 
\begin{equation}
S(t)={\cal{T}}exp\left\{-i\int^{t}_{0}H(t')dt' \right\}.
\label{Eq:Smatrix}
\end{equation}
Here ${\cal{T}}$ means temporal ordering
Te oscillation probability follows from 
\begin{equation}
 P (\nu_{\alpha}\rightarrow \nu_{\beta})\equiv P_{\nu_{\alpha}\rightarrow \nu_{\beta}}=|S_{\beta\alpha}|^{2}.
\end{equation}

From Eq.~(\ref{Eq:Hlinha}) we define the potential $\widetilde{V}=\widetilde{H}-\widetilde{H}^{(0)}$, which in the interaction picture assumes the form,
\begin{equation}
\widetilde{V}_{I}=e^{i\widetilde{H}^{(0)}t}\widetilde{V}e^{-i\widetilde{H}^{(0)}t},
\label{Eq:Vint}
\end{equation}
where $\widetilde{V}$ is the time-dependent potential in the Schroedinger picture. 
The time-evolution operator in the interaction picture can be defined as \cite{sakurai1995modern} $\Omega(x)$, 
\begin{equation}
\Omega(t)=e^{i \widetilde{H}^{(0)}t }\widetilde{S}(t),
\label{Eq:omega1}
\end{equation}
which  must obey the operator time evolution equation:
\begin{equation}
i\dfrac{d}{dt}\Omega(t)=\widetilde{V}_{I}\Omega(t)~,
\label{Eq:SCHR-int}
\end{equation}
which is subjected to the initial condition $\Omega(t=0)1$. The Eq.~(\ref{Eq:SCHR-int}) plus the initial condition implies in the integral equation:
\begin{equation}
\Omega(t)=1-i\int^{t}_{0} \widetilde{V}_{I}(t')\Omega(t')dt'.
\label{Eq-int-omega}
\end{equation}
The recursive substitution of $\Omega(x)$ into Eq.~(\ref{Eq-int-omega}) leads to the Dyson Series, which gives the solution for Eq.~(\ref{Eq:SCHR-int}) as:

\begin{eqnarray}
\Omega(x)=1+(-i)\int^x_0\widetilde{V}_{I}(x')dx'+(-i)^2\int^x_0\widetilde{V}_{I}(x')dx'\int^{x'}_0\widetilde{V}_{I}(x'')dx'' \notag \\
+(-i)^3\int^x_0\widetilde{V}_{I}(x')dx'\int^{x'}_0\widetilde{V}_{I}(x'')dx''\int^{x''}_0\widetilde{V}_{I}(x''')dx'''+\mathcal{O}(\epsilon^{2}).
\label{eq:Dyson}
\end{eqnarray}

Once $\Omega(x)$ is determinate from Eq.~(\ref{eq:Dyson}), in the propagation basis
\begin{eqnarray}
\widetilde{S}(x)=e^{-i\widetilde{H}^{(0)}}\Omega(x).
\label{eq:Dyson1}
\end{eqnarray}
The $S$ matrix 
can  then be written in the  flavor basis as 
\begin{equation}
S=[{\rm R} (\theta_{23})]^{\dagger}  \widetilde{S} {\rm R} (\theta_{23}) .
\label{S:flavor}
\end{equation}
The $\widetilde{S}$ matrix is  explicitly calculated in~\cite{Marianodissert}.

\section{Muon neutrino survival probability} 
\label{Ap:Pmm}

Here we show our results for muon neutrino survival probability obtained from the same procedure explained in Section \ref{Sec:Smat}. We consider terms until  $n \le 3/2$. 
The resulting formulas show the same functional structure of standard oscillation formalism, and the NSI features are incorporated in the $(\Sigma, \Omega, \Lambda, \Gamma)$ quantities,

\begin{eqnarray}
P_{\nu_{\mu}\nu_{\mu}}^{(0)}&=&1-4 c_{23}^2 s_{23}^2 \sin^2\left(\frac{\Delta_{31} x  r_A\Lambda}{2} \right),
\label{eq:Puu0}
\end{eqnarray}
\begin{eqnarray}
P_{\nu_{\mu}\nu_{\mu}}^{(1)}&=&-\frac{4 |\Sigma| ^2 s_{23}^4 }{r_A^2 (\Gamma -\Lambda )^2}\sin ^2\left(\frac{\Delta_{31} x  r_A (\Gamma -\Lambda )}{2} \right)\nonumber\\
&+&2 c_{23}^2  s_{23}^2\left(c_{12}^2 r_\Delta +\frac{|\Sigma| ^2}{r_A (\Gamma -\Lambda )}+s_{13}^2\right)\sin (\Delta_{31} x r_A\Lambda    ) (\Delta_{31} x)\nonumber\\
&+&\frac{2  |\Sigma| ^2 s_{23}^2 c_{23}^2 }{r_A^2 (\Gamma -\Lambda )^2}\left[\cos (\Delta_{31} x\, r_A\Gamma )-\cos (\Delta_{31} x\, r_A\Lambda )\right]\nonumber\\
&-&\frac{8  |\widetilde{\epsilon}_{\mu \tau}| c_{23} s_{23} \left(c_{23}^2-s_{23}^2\right) \cos (\widetilde{\phi}_{ \mu \tau }) }{\Lambda }\sin ^2\left(\frac{\Delta_{31} x \Lambda  r_A}{2}    \right),
\label{eq:Puu1}
\end{eqnarray}
\begin{eqnarray}
P^{(3/2)}_{\nu_{\mu}\nu_{\mu}}&=&-\frac{8  r_\Delta c_{12} s_{12}c_{23}  s_{23} s_{13}  \left(c_{23}^2-s_{23}^2\right) \cos (\delta_{\rm CP} ) }{r_A\Lambda  } \sin ^2\left(\frac{\Delta_{31} x\,  r_A \Lambda }{2}   \right)\nonumber \\
&+&\frac{4 |\Omega \Sigma| c_{23}  s_{23}  \cos \left(\phi _{\Sigma }-\phi _{\Omega }\right) }{  r_A^2 \Gamma(\Gamma -\Lambda )} \left(c_{23}^2 \cos (\Delta_{31} x\, r_A\Gamma   )-c_{23}^2+s_{23}^2\right) \cos  (\Delta_{31} x\, r_A  (\Gamma -\Lambda ) )\nonumber\\
&-&\frac{4 |\Omega \Sigma| c_{23}   s_{23}   \cos \left(\phi _{\Sigma }-\phi _{\Omega }\right) }{\Lambda  r_A^2 (\Gamma -\Lambda )}\left(c_{23}^2 \cos (\Delta_{31}   x\, r_A\Lambda  )+1\right)\nonumber\\
&+&\frac{4|\Omega\Sigma| c_{23}   s_{23}   \cos \left(\phi _{\Sigma }-\phi _{\Omega }\right) }{r_A^2\Gamma  \Lambda  }\left(s_{23}^2 \cos \Delta_{31}   x\, r_A\Lambda  )\right).
\end{eqnarray}
The muon neutrino probability it is 
\begin{eqnarray}
P^{\rm perturbative}(\nu_{\mu}\nu_{\mu})=P (\nu_{\mu}\nu_{\mu})^{(0)}+P( \nu_{\mu}\nu_{\mu})^{(1)}+P (\nu_{\mu}\nu_{\mu})^{(3/2)}.
\label{eq:Puuall}
\end{eqnarray}
\section{The Non-Standard Interaction Degeneracy}
\label{Ap:Deg}

We will discuss the conditions to have the standard oscillation probability to exactly degenerate with NSI oscillation probability.  

Here, from Eq.~\eqref{Eq:conddeg}, we will derive Eq.~\eqref{Eq:Eet2}. Substituting the explicit expression for the probability, Eq.~\eqref{eq:result1}, on the expression for the degeneracy condition, Eq.~\eqref{Eq:conddeg}, it give us

\begin{eqnarray}
\left[ P^{(1)} (\nu_{\mu}\to\nu_{e})(\widetilde{\epsilon}_{\alpha\beta},\pmb{\theta_{23}})\right]^{\mbox{(\rm NSI)}}=
\left[ P^{(1)} (\nu_{\mu}\to \nu_{e}) (\widetilde{\epsilon}_{\alpha\beta}=0,\theta_{23})\right]^{\mbox{(\rm SO)}} & &    \nonumber \\
    4\frac{ \left|\Sigma\right|^2 (\pmb{s_{23}})^2 }{r_A^2 \eta^2}\sin^{2} \left(\frac{\Delta_{31} x}{2} r_A \eta \right)=4\frac{ s_{13}^2 s_{23}^2 }{(1-r_A)^2}\sin^{2} \left(\frac{\Delta_{31} x}{2} (1-r_A) \right),
\end{eqnarray}
that can be written as

\begin{equation}
\frac{r^{2}_{A}}{s_{13}^2}|\widetilde{\epsilon}_{e\tau}|^{2}+2\frac{r_{A}}{s_{13}}|\widetilde{\epsilon}_{e\tau}|\cos(\zeta)+1=\left(\frac{s_{23}}{\pmb{s_{23}}}\right)^2\frac{\text{sinc}^2(\Delta_{31} x(1-r_A))}{\text{sinc}^2(\Delta_{31} x\eta)},
\label{Eq:rsinc1}
\end{equation}
 where the functions $\text{sinc}(x)$ are defined as $\text{sinc}(x)\equiv \frac{sin(x)}{x}$. We can expand the $\text{sinc}(\Delta_{31} x\eta)$ around the standard oscillation phase, and it will give us
 
 \begin{eqnarray}
 & &\frac{\text{sinc}^2(\Delta_{31} x(1-r_A))}{\text{sinc}^2(\Delta_{31} x \eta)}\approx
\left(\frac{\text{sinc}(\Delta_{31} x(1-r_A))}{\text{sinc}(\Delta_{31} x(1-r_A))+B_2r_A(\widetilde{\epsilon}_{ee}-\widetilde{\epsilon}_{\tau \tau})}\right)^2,
\label{Eq:rsincapprox1}
\end{eqnarray}
where  can define the quantities
\begin{eqnarray}
B_1&=&\text{sinc}\left(\frac{\Delta_{31} x}{2}(1-r_{A})\right),\nonumber \\
B_2&=&\left(\frac{\Delta_{31} x}{2}\right)\frac{\cos\left(\frac{\Delta_{31} x}{2}(1-r_{A})\right)-\text{sinc}\left(\frac{\Delta_{31} x}{2}(1-r_{A})\right)}{\left(\frac{\Delta_{31} x}{2}(1-r_{A})\right)}.\nonumber\\
\label{Eq:B1B21}
\end{eqnarray}
Replacing Eq.~(\ref{Eq:rsincapprox1}) into 
Eq.~(\ref{Eq:rsinc1}) we have
\begin{equation}
\frac{r^{2}_{A}}{s_{13}^2}|\widetilde{\epsilon}_{e\tau}|^{2}+2\frac{r_{A}}{s_{13}}|\widetilde{\epsilon}_{e\tau}|\cos(\zeta)+1
=\left(\frac{s_{23}}{\pmb{s_{23}}}\right)^2\left(\frac{B_1}{B_1+B_2r_A(\widetilde{\epsilon}_{ee}-\widetilde{\epsilon}_{\tau\tau})}\right)^2.
\end{equation}
where this relation give the condition for the equality of neutrino probabilities for a given set of NSI parameters $\widetilde{\epsilon}_{ee},\widetilde{\epsilon}_{\tau\tau},\widetilde{\epsilon}_{e\tau},(\zeta$ and the mixing angle $\theta_{23}$).

\end{document}